\documentclass[11pt]{scrartcl}
\pdfoutput=1
\usepackage{amsmath,amssymb}
\usepackage{graphicx}
\usepackage{subfigure}
\usepackage{hyperref}
\usepackage{nicefrac}
\usepackage{color}
\usepackage{tikz}

\newcommand{\re}{\mathrm{Re}}

\graphicspath{{plots/}}

\begin{document}

\title{Hadron-Hadron Interactions from $N_f=2+1+1$ Lattice QCD:\\
  isospin-2 $\pi\pi$ scattering length}

\author{C.~Helmes, C.~Jost, B.~Knippschild, L.~Liu, C.~Urbach,\\
  M.~Ueding and M.~Werner\\
  \small{Helmholtz Institut f{\"u}r Strahlen- und Kernphysik, University of Bonn, Bonn, Germany}\vspace*{0.3cm}\\
  \and
  C.~Liu\\
  \small{School of Physics and Center for High Energy Physics, Peking
    University, Beijing, China}\\
  \small{Collaborative Innovation Center of Quantum Matter, Beijing, China}\vspace*{0.3cm}\\ 
  \and
  J.~Liu and Z.~Wang\\
  \small{School of Physics, Peking University, Beijing, China}
} 

\maketitle

\begin{abstract}
  \begin{center}
    \includegraphics[draft=false,width=.2\linewidth]{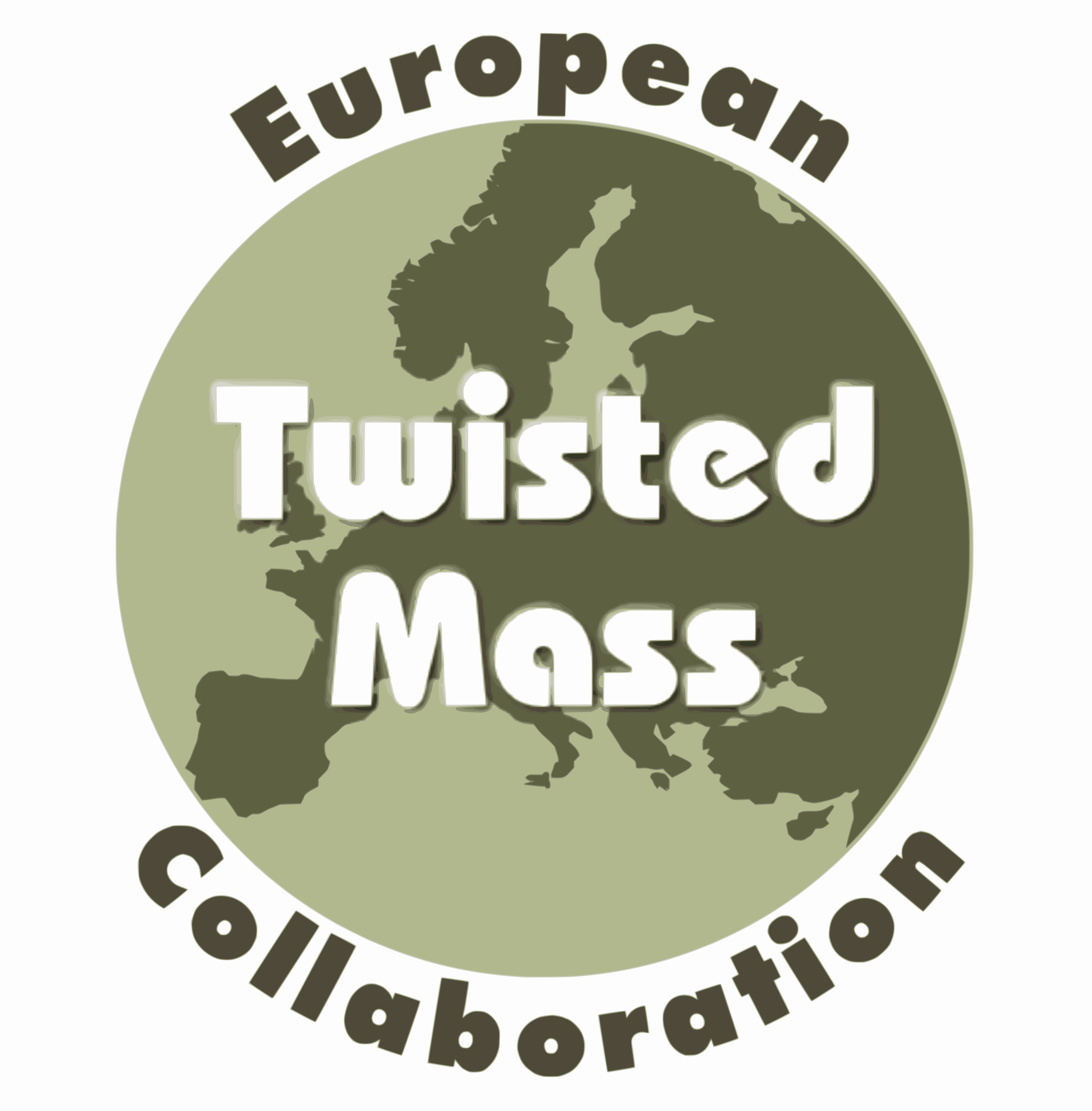}
  \end{center}
  \vspace*{0.3cm}
  We present results for the $I=2$ $\pi\pi$ scattering length using
  $N_f=2+1+1$ twisted mass lattice QCD for three values of the lattice
  spacing and a range of pion mass values. Due to the use of Laplacian
  Heaviside smearing our statistical errors are reduced compared to
  previous lattice studies. A detailed investigation of systematic
  effects such as discretisation effects, volume effects, and pollution
  of excited and thermal states is performed. After extrapolation to
  the physical point using chiral perturbation theory at NLO we obtain
  $M_\pi a_0=-0.0442(2)_\mathrm{stat}(^{+4}_{-0})_\mathrm{sys}$.
\end{abstract}

\clearpage

\section{Introduction}

Quantum Chromodynamics (QCD) describes, beyond the mass spectrum of stable
and unstable hadrons, also their interaction due to the strong force. While originally thought
to be not accessible to lattice QCD~\cite{Maiani:1990ca}, Martin
L{\"u}scher was able to relate the energy spectrum in finite volume to the
infinite volume interaction properties in a series of seminal
papers~\cite{Luscher:1985dn,Luscher:1986pf,Luscher:1990ck,Luscher:1990ux}.
During the last years, extensions of these finite volume techniques
have been worked out and they have been applied to a number of
interesting systems.

L{\"u}scher's original formalism dealt with two massive bosons in
a finite cubic box below the inelastic scattering threshold. It was
originally formulated for the center of mass frame.
In order to enlarge the applicability of L\"uscher's method, 
extensions have been developed over the years which
include: moving frames with non-vanishing center of mass
momentum~\cite{Rummukainen:1995vs,Kim:2005gf,Feng:2011ah,Gockeler:2012yj},
asymmetric boxes~\cite{Li:2003jn,Feng:2004ua,Detmold:2004qn}, twisted
boundary
conditions~\cite{Bedaque:2006yi,Doring:2011vk,Briceno:2013hya,Agadjanov:2013kja},
 and more. Most of these aim to circumvent the poor resolution
in momentum space due to the quantisation of the
three-momentum in finite volume. Generalisations beyond the inelastic threshold for the
two particle cases are also
discussed~\cite{He:2005ey,Liu:2005kr,Bernard:2010fp,Hansen:2012tf,Briceno:2012yi,Guo:2012hv}
and various groups are now working on the more difficult
three-particle
scenario~\cite{Roca:2012rx,Polejaeva:2012ut,Briceno:2012rv,Hansen:2013dla,Hansen:2015zga,Hansen:2014lya,Hansen:2014eka}.

In this first paper of a planned series of papers we investigate
$\pi\pi$ scattering in the isospin-2 channel. $\pi\pi$ scattering in
the $I=2$ channel is technically the easiest 
case to consider in lattice QCD, because so-called fermionic
disconnected contributions are absent. Moreover, the predicted quark
mass dependence in chiral perturbation theory ($\chi$PT) at
next-to-leading order (NLO) is governed by only one low energy constant, and
so far lattice data showed surprisingly little deviations even from
the leading order (LO) $\chi$PT predictions, which is parameter
free. Isospin-2 $\pi\pi$ scattering is, therefore, an important
benchmark system to compare with non-lattice methods and an
important test case for future investigations of hadron-hadron
interactions, while simultaneously of phenomenological interest.

Consequently, it has been computed in lattice QCD
previously~\cite{Beane:2007xs,Feng:2009ij,Dudek:2010ew,Beane:2011sc,Dudek:2012gj}
using $N_f=2$ or $N_f=2+1$ dynamical fermions, for a recent review see
Ref.~\cite{Prelovsek:2014zga}. In this paper we extend
this list by using $N_f=2+1+1$ dynamical quark
flavours for the first time. The analysis relies on Wilson twisted mass
fermions~\cite{Frezzotti:2000nk} at maximal
twist~\cite{Frezzotti:2003ni} based on gauge
configurations provided by the European Twisted Mass Collaboration
(ETMC)~\cite{Baron:2010bv,Baron:2010th}. This allows us to investigate
a range of pion masses from $250$ to $500\
\mathrm{MeV}$ and discretisation effects by using 
three values of the lattice spacing. For estimating finite volume
effects we have several ensembles at our disposal with all parameters
fixed but the volume.

We apply different strategies to determine scattering parameters from
the data using L{\"u}scher's finite volume method. This gives us an
estimate of residual finite volume effects on the scattering length
for other ensembles where we do not have multiple volumes at hand.

The main result is an extrapolation of $M_\pi a_0$ to the physical
point utilising chiral perturbation theory at next-to-leading
order. We carry systematic effects from different fit ranges through
the entire analysis chain. In addition we investigate the stability of
the extrapolation by different cuts in the pion mass values. 

The paper is structured as follows: in section~\ref{sec:actions} we review
the fermion action, the Laplacian-Heaviside smearing technique and
the interpolating operators we use to construct the correlation
functions. L{\"u}scher's finite volume method is explained in
section~\ref{sec:fs}. Our analysis and 
the treatment of all sources of systematic effects are worked out in
section~\ref{sec:results}. The final results are discussed in
section~\ref{sec:discussion} and summarised in
section~\ref{sec:summary}.

\section{Lattice action}
\label{sec:actions}

\begin{table}[t!]
 \centering
 \begin{tabular*}{.9\textwidth}{@{\extracolsep{\fill}}lcccccc}
  \hline\hline
  ensemble & $\beta$ & $a\mu_\ell$ & $a\mu_\sigma$ & $a\mu_\delta$ &
  $(L/a)^3\times T/a$ & $N_\mathrm{conf}$  \\ 
  \hline\hline
  $A30.32$   & $1.90$ & $0.0030$ & $0.150$  & $0.190$  &
  $32^3\times64$ & $280$  \\
  $A40.20$   & $1.90$ & $0.0040$ & $0.150$  & $0.190$  &
  $20^3\times48$ & $553$  \\
  $A40.24$   & $1.90$ & $0.0040$ & $0.150$  & $0.190$  &
  $24^3\times48$ & $404$  \\
  $A40.32$   & $1.90$ & $0.0040$ & $0.150$  & $0.190$  &
  $32^3\times64$ & $250$  \\
  $A60.24$   & $1.90$ & $0.0060$ & $0.150$  & $0.190$  &
  $24^3\times48$ & $314$  \\
  $A80.24$   & $1.90$ & $0.0080$ & $0.150$  & $0.190$  &
  $24^3\times48$ & $306$  \\
  $A100.24$  & $1.90$ & $0.0100$ & $0.150$  & $0.190$  &
  $24^3\times48$ & $312$  \\
  \hline
  $B35.32$   & $1.95$ & $0.0035$ & $0.135$  & $0.170$  &
  $32^3\times64$ & $250$ \\
  $B55.32$   & $1.95$ & $0.0055$ & $0.135$  & $0.170$  &
  $32^3\times64$ & $311$ \\
  $B85.24$   & $1.95$ & $0.0085$ & $0.135$  & $0.170$  &
  $32^3\times64$ & $296$ \\
  \hline
  $D45.32sc$ & $2.10$ & $0.0045$ & $0.0937$ & $0.1077$ &
  $32^3\times64$ & $301$ \\
  \hline\hline
 \end{tabular*}
 \caption{The gauge ensembles used in this study. For the labelling of
   the ensembles we adopted the notation in
   Ref.~\cite{Baron:2010bv}. In addition to the relevant input
   parameters we give the lattice volume and  the number of evaluated
   configurations, $N_\mathrm{conf}$.}
 \label{tab:setup}
\end{table}

The sea quarks are described by the Wilson twisted mass action with
$N_f=2+1+1$ dynamical quark flavours. The Dirac operator for the light
quark doublet reads~\cite{Frezzotti:2000nk} 
\begin{equation}
  D_\ell = D_\mathrm{W} + m_0 + i \mu_\ell \gamma_5\tau^3\, ,
  \label{eq:Dlight}
\end{equation}
where $D_\mathrm{W}$  denotes the standard Wilson Dirac operator and $\mu_\ell$
the bare light twisted mass parameter. $\tau^3$ and in general
$\tau^i, i=1,2,3$ represent the Pauli matrices acting in flavour
space. $D_\ell$ acts on a spinor $\chi_\ell = (u,d)^T$ and, hence, the $u$
($d$) quark has twisted mass $+\mu_\ell$ ($-\mu_\ell$).

For the heavy unitary doublet of $c$ and
$s$ quarks~\cite{Frezzotti:2003xj} the Dirac operator is given by
\begin{equation}
  D_\mathrm{h} = D_\mathrm{W} + m_0 + i \mu_\sigma \gamma_5\tau^1 + \mu_\delta \tau^3\,.
  \label{eq:Dsc}
\end{equation}
The bare Wilson quark mass $m_0$ has been tuned to its critical
value~\cite{Chiarappa:2006ae,Baron:2010bv}. This guarantees
automatic order $\mathcal{O}\left(a\right)$ improvement
\cite{Frezzotti:2003ni}, which is one of the main advantages of the
Wilson twisted mass formulation of lattice QCD. 

The splitting term in the heavy doublet Eq.~\ref{eq:Dsc} introduces flavour mixing
between strange and charm quarks which needs to be accounted for in
the analysis. However, this is only important for quantities involving valence strange
and charm quarks.

%All errors are computed using a blocked bootstrap procedure
%to account for autocorrelation. The number of bootstrap samples was
%taken to be $1000$ and the number of configurations per block $N_b$ is
%given for every ensemble in table~\ref{tab:setup}. $N_b$ itself was
%chosen such that the length of a block corresponds to at least $20$
%HMC trajectories of length one. This value turned out sufficient to
%compensate for autocorrelation in the observables considered in this
%study.

\subsection{Stochastic Laplacian Heavyside Smearing}

On the long term we plan to address as many observables as possible. Hence
we have decided to use the stochastic Laplacian Heavyside smearing (sLapH),
which is quite generally applicable and which is described in detail
in Refs.~\cite{Peardon:2009gh,Morningstar:2011ka}. It represents a smearing method based 
on the covariant 3-dimensional Laplace operator 
\begin{equation}
   \widetilde{\Delta}^{ab}(x,y;U) = \sum_{k=1}^3 \Bigl\{
   \widetilde{U}_k^{ab}(x)\delta(y,x+\hat{k})  + \widetilde{U}^{ba}_k(y)^\dagger\delta(y,x-\hat{k})
   - 2\delta(x,y)\delta^{ab}\Bigr\} \,,      
  \label{eq:laplace}
\end{equation}
where $a, b$ are colour indices, $x, y$ space-time coordinates,
$\widetilde U$ denote smeared gauge links, and the sum is taken over
the three spatial directions. The gauge fields in the Laplace
operator, Eq.~\ref{eq:laplace}, are smeared with three steps of three
dimensional HYP smearing \cite{Hasenfratz:2001hp} with smearing
parameters $\alpha_1=\alpha_2=0.62$, independently of the volume and
lattice spacing. The Laplace operator can be 
decomposed as follows
\begin{equation}
  \widetilde\Delta = V_\Delta \Lambda_\Delta V_\Delta^\dagger\,,
\end{equation}
where $V_\Delta$ represents the matrix of all eigenvectors and
$\Lambda_\Delta$ is a diagonal matrix containing the eigenvalues.  
Colour, Dirac and space-time indices are suppressed.
The smearing matrix then reads
\begin{equation}
  \mathcal{S} = V_\mathrm{s} V_\mathrm{s}^\dagger\,,
\end{equation}
where $V_\mathrm{s}$ only contains eigenvectors corresponding to eigenvalues
smaller than a cut-off $\sigma_s^2$. These eigenvectors span what is
called the LapH sub-space. The value $\sigma_s^2$ can be chosen in a
way that  excited state contaminations are maximally suppressed.
In Ref.~\cite{Morningstar:2011ka} it has been found that $\sigma_s^2=0.33$ is
optimal for excited state suppression independent of the interpolating
operator. On a $48^3\times96$ lattice, for example, this amounts to
about 960 eigenvectors per time slice. However, our tests show that it
is better to have less eigenvectors for these large lattices, namely
$660$ on $48^3\times96$ and $220$ on $32^3\times64$, to suppress
excited state contaminations at early Euclidean times.

The building blocks of observables are quark lines, $\mathcal{Q}$,
which can be written in the LapH framework as 
\begin{equation}
  \mathcal{Q} = \mathcal{S}\Omega^{-1}\mathcal{S} =
  V_\mathrm{s}\ (V_\mathrm{s}^\dagger \Omega^{-1} V_\mathrm{s}) 
  \ V_\mathrm{s}^\dagger \,,
  \label{eq_quarklie}
\end{equation} 
where the middle part, $\mathcal{P} = V_\mathrm{s}^\dagger \Omega^{-1} V_\mathrm{s}$, is called perambulator with $\Omega =
\gamma_4 M$ and $M$ being the Dirac operator. 
To generate a perambulator it is necessary to invert the Dirac operator for every eigenvector in $V_\mathrm{s}$. 
It follows that for an all-to-all perambulator, since the Laplace operator is diagonal in time and Dirac space, 
this procedure needs to be repeated for each time slice and Dirac component. 
Taking the $48^3\times96$ lattice as an example this would result in a total number of $253440$ inversions. 
This number can be reduced significantly by combining the LapH method with a stochastic approach as described
in Ref.~\cite{Morningstar:2011ka} in detail.

For constructing all-to-all propagators with a stochastic approach,
random vectors $\rho$ are introduced which carry Dirac, time, colour
and spatial indices. A propagator can be computed by solving 
\begin{equation}
	\Omega X^{r[b]} = \rho^{r[b]}
	\label{eq_random_prop}
\end{equation}
for $X^{r[b]}$. Here the random vectors are
not chosen to be random valued in all elements, but diluted, which is
used to reduce the stochastic noise, see
Ref.~\cite{Morningstar:2011ka} for details. Correspondingly, the
composed index $r[b]$ counts the total number of random vectors,
$N_\mathrm{R}$, via $r$ and the total number of dilution vectors, $N_\mathrm{D}$, via
$b$. When constructing the all-to-all propagator via 
\begin{equation}
   \Omega^{-1}\approx \frac{1}{N_\mathrm{R}}
 \sum_{r=1}^{N_\mathrm{R}} \sum_b X^{r[b]}\rho^{r[b]\dagger}\,,
\end{equation}
the zeros in the diluted random vectors ensure exact zeros in the
product $\rho^{r[b]}\rho^{r[b]\dagger}$, which reduces noise
significantly. However, diluting 
random vectors leads to higher computational costs due to solving
Eq.~\ref{eq_random_prop} for each of these vectors separately. It is expected
that the noise in correlation functions built from diluted
stochastic propagators reduces with $\nicefrac{1}{\sqrt{N_\mathrm{R}}}$ and
$\nicefrac{1}{N_\mathrm{D}}$, which favours more dilution vectors over more
random vectors. Aside from this, each quark line needs its own set of
random vectors to avoid a bias in the correlation functions. 

In the stochastic version of LapH, denoted by sLapH, the random
vectors are introduced in time, Dirac and Laph sub-space indices. 
A quark line can then be estimated stochastically via
\begin{equation}
 {\mathcal Q} \approx \frac{1}{N_\mathrm{R}}\sum_{r=1}^{N_\mathrm{R}}\sum_b  
  {\mathcal S} \Omega^{-1}\ V_\mathrm{s} \rho^{r[b]}
  \  (V_\mathrm{s} \rho^{r[b]})^\dagger\,.
  \label{eq_stoch_quark_line}
\end{equation}
In our implementation we choose $Z_2$ random numbers. As mentioned
before each quark line as defined in Eq.~\ref{eq_stoch_quark_line},
needs its own random vector to avoid a bias. This means that at least
four random vectors are needed to be able to compute the correlation
functions relevant for $\pi\pi$ scattering processes. However, we use 
five random vectors, since the fifth random vector will allow for
additional permutations in the four point functions. This improves the
signal-to-noise ratio by a factor of 2 by only increasing the computing
costs for inversions by $25\%$.  

As dilution scheme we have chosen full dilution in Dirac space combined
with a block dilution in time and an interlace dilution in the LapH
sub-space. The number of dilution vectors are summarised in
table~\ref{table_dilutionvectors}. They were chosen in such a way that
the number of inversions and the noise in our observables are minimised
simultaneously.  

We remark here that with the sLapH smearing scheme as explained above,
only smeared-smeared correlation functions can be computed. Hence, we
cannot compute matrix elements of local operators needed for instance
for $f_\pi$ without major additional effort.

\begin{table}[h]
  \centering
  \begin{tabular*}{.9\textwidth}{@{\extracolsep{\fill}}crrrr}
    \hline\hline
    $(L/a)^3\times T/a$     & $N_\mathrm{D}(\text{time})$ & $N_\mathrm{D}(\text{Dirac})$ &
    $N_\mathrm{D}(\text{LapH})$ & total \# inversions \\
    \hline\hline
    $20^3\times48$  & 24    & 4  & 6  & $5\cdot576=2880$   \\
    $24^3\times48$  & 24    & 4  & 6  & $5\cdot576=2880$   \\
    $32^3\times64$  & 32    & 4  & 4  & $5\cdot512=2560$   \\
    $48^3\times96$  & 32    & 4  & 4  & $5\cdot512=2560$   \\
    \hline\hline
  \end{tabular*}
  \label{table_dilutionvectors}
  \caption{Summary of the number of dilution vectors, $N_\mathrm{D}$, used in
    each index. We use a block scheme in time and an interlace scheme
    in eigenvector space. The total number of inversions is the number
    of random vectors, here 5, multiplied by the number of inversions
    for one quark line.}
\end{table}

\subsection{Operators}

The phase shifts are extracted from the finite-volume energies as
described in the next section. In order to estimate the finite-volume
energy spectrum, we build a matrix of correlators with a set of
operators that resemble the $\pi\pi$ isospin-2 system, and then use
the variational method~\cite{Michael:1982gb,Luscher:1990ck} to extract
the spectral information. 

The $SO(3)$ symmetry in continuum space is reduced to the octahedral
group $O$ on the lattice. For non-zero momentum $\vec{P}$ the symmetry
is further reduced to the little group $LG(\vec{P})$ that leaves
$\vec{P}$ invariant. We construct the operators with definite total
momentum $\vec{P}$ in each irreducible representation (irrep) of
$LG(\vec{P})$ via:
\begin{equation}
  \mathcal{O}^{\vec{P}, \Lambda, \lambda}_{\vec{p}_1, \vec{p}_2} ( t )
  = \sum_{\substack{\vec{p}_1+\vec{p}_2 = \vec{P} \\ \vec{p}_1\in
      \{\vec{p}_1\}^* \\ \vec{p}_2\in \{\vec{p}_2\}^* }}
  \mathcal{C}(\vec{P}, \Lambda, \lambda; \vec{p}_1, \vec{p}_2) \,
  \pi^+ (\vec{p}_1, t) \pi^+ (\vec{p}_2, t+1)\,. 
\end{equation}
Here, $\pi^+(\vec{p}, t) = \sum_{\vec{x}} e^{i\vec{p} \cdot \vec{x}}
\bar{d}(\vec{x}, t) \gamma_5 u(\vec{x}, t)$, is the single pion
operator projected onto momentum $\vec{p}$. With periodic boundary
conditions, $\vec{p}$ is quantised as $\vec{p} = \frac{2\pi}{L}
\vec{n}$, where $\vec{n}$ is a vector of integers.  $\Lambda$ is an
irrep of the group $LG(\vec{P})$ and $\lambda$ is the irrep row.
$\{\vec{p}_{1,2}\}^*$ represents the set of vectors $\{R \,
\vec{p}_{1,2}, R \in O \}$. $ \mathcal{C}(\vec{P}, \Lambda, \lambda;
\vec{p}_1, \vec{p}_2)$ are the Clebsch-Gordan coefficient for
$\Lambda_1 \otimes \Lambda_2 \to \Lambda$, where
$\Lambda_1$($\Lambda_2$) is the irrep that
$\pi^+(\vec{p}_1)$($\pi^+(\vec{p}_2$)) resides in.  
The two pions are separated by one time slice in order to avoid the
complications due to Fierz rearrangement~\cite{Fukugita:1994ve}.

In this work we focus on building the operators for total zero
momentum $\vec{P} = [0, 0, 0]$ and $A_1(\ell = 0, 4, \dots)$ irrep where
we can safely ignore all partial waves higher than $\ell=0$ in our
analysis. The Clebsch-Gordan coefficients $\mathcal{C}(\vec{P},
\Lambda, \lambda; \vec{p}_1, \vec{p}_2)$ are taken from
Ref.~\cite{Dudek:2012gj}.

The correlation matrix is computed via:
\begin{equation}
  \label{eq:corM}
  C_{ij}^{\vec{0}, \Lambda} (t) = \langle 0 | \mathcal{O}_i^{\vec{0},
    \Lambda}( t ) \, {\mathcal{O}_j^{\vec{0}, \Lambda} (0)}^\dagger |
  0 \rangle\,. 
\end{equation}
The variational basis contains various combinations of $\vec{p}_1$ and
$\vec{p}_2$ that are allowed by the decomposition $\Lambda_1 \otimes
\Lambda_2 \to \Lambda$. For the case of $\vec{P} = [0, 0, 0]$ we
include for the  
$A_1$ irrep the operators with $|\vec{p}_{1,2} | = (0, 1, 2,
3, 4)$, which gives a 5-dimensional correlation matrix. More details
about the irreps and other $\vec{P}$ values can be found in
Ref.~\cite{Dudek:2012gj}.

The energies are obtained by solving the generalised eigenvalue problem: 
\begin{equation}
  C(t) v_i(t, t_0) = \lambda_i(t, t_0) C(t_0) v_i(t, t_0)\,.
\end{equation}
It can be shown that the eigenvalues $\lambda_i(t)$ behave like
\begin{equation}
  \lambda_i (t) \sim e^{-E_i (t - t_0)} + \cdots\,,
\end{equation}
where $E_i$ is the $i$-th eigenvalue of the Hamiltonian of the
system. However, we focus on zero total momentum where apparently
solving the GEVP does not give any advantage over using $\vec
p_1=\vec p_2=\vec 0$ only. This is due to a rather weak coupling of
different momenta in the matrix of correlators. All results presented in the
following are, therefore, obtained directly from the four-point
function at $\vec p_1=\vec p_2=\vec 0$. 

\subsection{Removing Thermal States}
\label{sec:temporal}

As discussed in Ref.~\cite{Feng:2009ij} and references therein, the
spectral analysis of the two pion correlation function in the case of
total zero momentum deviates from the usual $\cosh$ like behaviour. 
It was shown in Ref.~\cite{Feng:2009ij} that the diagonal
elements of the correlation matrix, Eq.~\ref{eq:corM}, for the two
particle system obey the following spectral decomposition in
the limit of large Euclidean times
\begin{equation}
  \label{eq:Tconst}
  C_{\pi\pi}(t)\equiv C_{00}^{\vec 0, \Lambda}(t)\ \propto A_0
  \cosh(E_{\pi\pi}(t-T/2)) + c \exp(-M_\pi T)\,, 
\end{equation}
with $E_{\pi\pi}$ the two pion energy and $M_\pi$ the single pion mass,
respectively, and constants $A_0$ and $c$.
This spectral decomposition differs from the standard by the term constant in
Euclidean time. In the thermodynamic limit $T\to\infty$ this polluting
term vanishes, but for finite $T$ it will dominate the correlation
function at large Euclidean times. To remove this pollution, it was
proposed to take a finite difference first~\cite{Umeda:2007hy} and
then build the following ratio~\cite{Feng:2009ij}
\begin{equation}
  \label{eq:ratio}
  R(t+\nicefrac{1}{2})\ =\ \frac{C_{\pi\pi}(t) - C_{\pi\pi}(t+1)}{C_{\pi}^2(t) - C_\pi^2(t+1)}\,,
\end{equation}
with $C_\pi(t)$ the single pion two-point correlation function. One can show
that the ratio has the functional form~\cite{Feng:2009ij}
\begin{equation}
  \label{eq:ratio2}
  R(t+\nicefrac{1}{2}) = A(\cosh(\delta E\ t') + \sinh(\delta E\ t') \coth(2 E_\pi
  t'))
\end{equation}
with $t' = t+1/2-T/2$ and $\delta E = E_{\pi\pi} - 2M_\pi$ the energy shift.

The generalisation of this procedure to correlation matrices and the
variational method persists in shifting the correlation matrix
Eq.~\ref{eq:corM} before using the standard GEVP procedure, see
Ref.~\cite{Dudek:2012gj} for details.

\section{Finite Volume Methodology}
\label{sec:fs}

We are interested in the limit of small scattering momenta for the
$\pi\pi$ system with $I=2$ below inelastic threshold. Using the finite
range expansion, the scattering length 
$a_0$ and the effective range $r_0$ can be related to the energy shift
$\delta E$ by an expansion in $1/L$ as follows~\cite{Luscher:1986pf}
\begin{equation}
  \label{eq:luscher1}
  \delta E = - \frac{4 \pi a_0} {M_\pi L^3} \left(1 +c_1 \frac{a_0}{L}
  + c_2 \frac{a_0^2}{L^2} + c_3\frac{a_0^3}{L^3}\right) -
  \frac{8\pi^2 a_0^3}{M_\pi L^6}r_0 +
  \mathcal{O}(L^{-7})\,,
\end{equation}
with $\delta E = E_{\pi\pi} - 2M_\pi$ and coefficients~\cite{Luscher:1986pf,Beane:2007qr}
\[
c_1 =  -2.837297\,,\quad c_2=6.375183\,,\quad c_3=-8.311951\,. 
\]
More generally, including also non-zero total momentum, L{\"u}scher's
method relates the phase shifts $\delta$ to the finite volume energy
shift via the relation
\begin{equation}
  \label{eq:det}
  \det\left[ e^{2i\delta}(\mathbf{M}-i)-(\mathbf{M}+i)\right]\ =\ 0\,,
\end{equation}
where the matrix elements of the matrix $\mathbf{M}$ are given as~\cite{Rummukainen:1995vs}
\begin{equation}
  \label{eq:M}
  M_{lm,l'm'}^{\vec d}(q) =
  \gamma^{-1}\frac{(-1)^l}{\pi^{3/2}}\sum_{j=|l-l'|}^{l+l'}\sum_{s=-j}^{j}\frac{i^j}{\tilde
    q^{j+1}}\mathcal{Z}_{js}^{\vec{d}}(1,\tilde
  q^2)C_{lm,js,l'm'}\,.
\end{equation}
$\mathcal{Z}$ is L{\"u}scher's generalised $\mathcal{Z}$-function and 
\begin{equation}
  \label{eq:qtilde}
  \tilde q = \frac{qL}{2\pi}\,.
\end{equation}
the lattice scattering momentum. The elements of the tensor
$C_{lm,js,l'm'}$ can be found in Ref.~\cite{Rummukainen:1995vs}.
For the case $\ell=0$ with total zero momentum and no mixing with higher
partial waves, Eq.~\ref{eq:det} reduces to 
\begin{equation}
  \label{eq:qcotdelta}
  q \cot\delta_0\ =\ \frac{2}{L \sqrt{\pi}}\ \re\{
  \mathcal{Z}_{00}^{\vec d}(1,\tilde q^2)\} \,.
\end{equation}
The scattering momentum $q$ is then given as 
\begin{equation}
  \label{eq:qcont}
  q^2 = \frac{(E_{\pi\pi})^2}{4}-M_\pi^2\,,
\end{equation}
from which its lattice version, Eq.~\ref{eq:qtilde}, can be computed.

As proposed in Ref.~\cite{Rummukainen:1995vs} the continuum dispersion
relation can be replaced by a lattice modified one. However, we do not
see any difference in using one or the other version of the dispersion
relation for the case of zero total momentum studied here. Hence, we
stick to the continuum dispersion relation for this paper.

\section{Results}
\label{sec:results}

\begin{figure}[t]
  \centering
  \includegraphics[width=\linewidth]{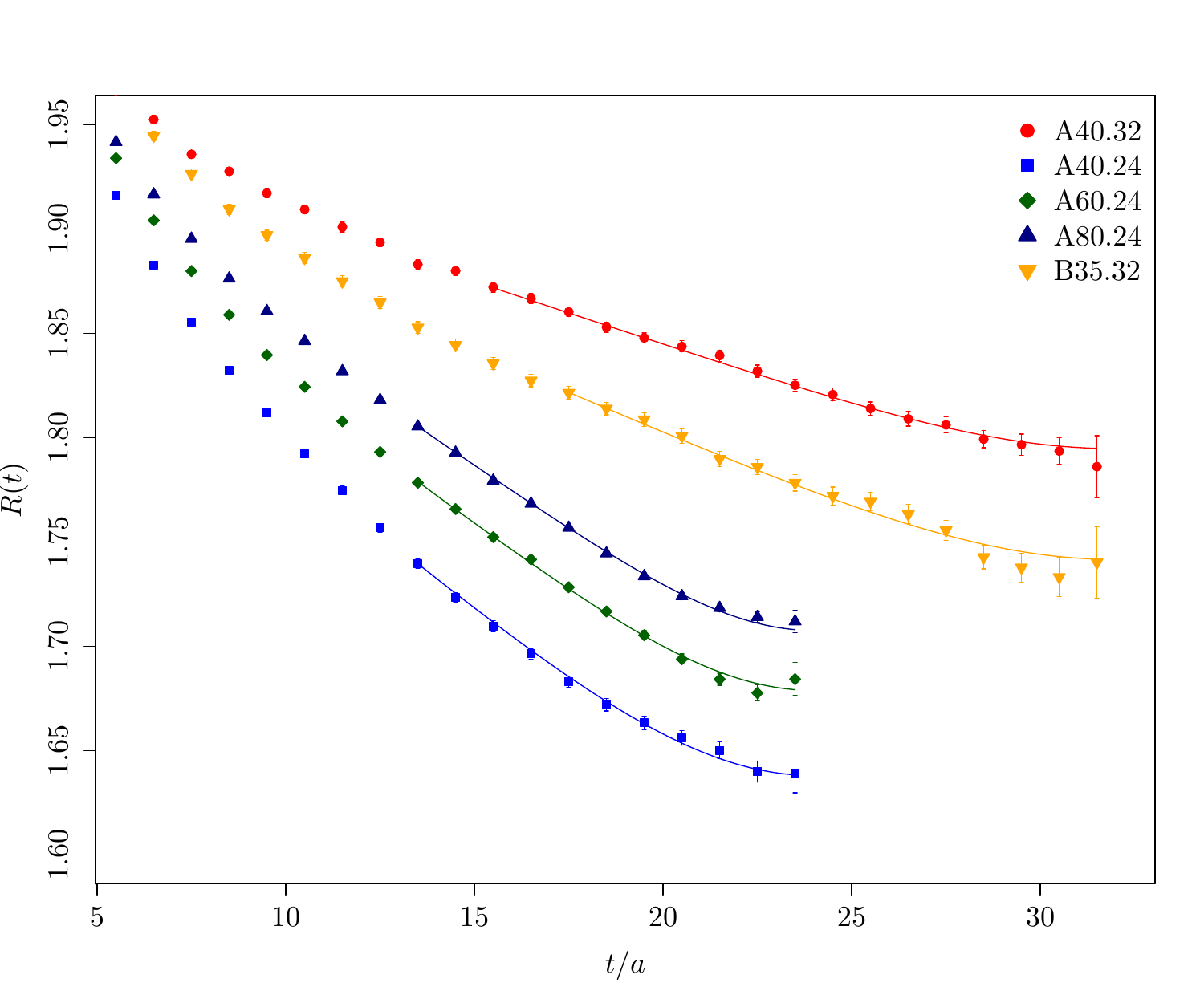}
  \caption{Ratios R(t) as defined in Eq.~\ref{eq:ratio} as a
    function of $t/a$ for different ensembles in the center of mass
    frame. We also show the best fit curves for some representatively
    chosen fit ranges.}
  \label{fig:ratios}
\end{figure}

Before coming to the actual results, let us first describe our
analysis procedure. Statistical errors are always computed using a
bootstrap procedure with $R=1500$ bootstrap samples. The bootstrap
analysis is chained such that also statistical errors for best fit
parameters can be determined. The gauge configurations are
sufficiently separated HMC trajectories that
autocorrelation does not play a role here, as we explicitly checked
using the blocked bootstrap procedure.

Energy shifts are determined from fully correlated fits of the ratios,
Eq.~\ref{eq:ratio} to the data. The single pion energy
levels needed for the fit are determined from the corresponding two point
functions using a one state exponential ($\cosh$) fit to the data,
again fully correlated. The correlation between the single pion
energy level and the ratio data is taken into account by using the
same bootstrap samples. 

In figure~\ref{fig:ratios} we
show example plots of the ratio for various ensembles. 
For a representative choice of the fit range we 
also show the corresponding best fit curves obtained by fitting
Eq.~\ref{eq:ratio2} to the data. For the fit ranges shown one observes
visually a good agreement between fitted curve and data.

The fits to the ratios and the two point
functions are repeated for a large number of fit ranges. We then
assign a weight 
\begin{equation}
  \label{eq:weight}
  w_\mathrm{X} = \{(1-2|p_\mathrm{X} - 0.5|) \cdot\min(\Delta_\mathrm{X})/\Delta_\mathrm{X}\}^2\,,  
\end{equation}
to every of these fits and quantities $X=E_\pi, E_{\pi\pi}$, where
$p_\mathrm{X}$ is the p-value of the corresponding fit and $\Delta_\mathrm{X}$ the
statistical error of $\langle X\rangle$ determined from the bootstrap
procedure. 

\begin{figure}[h!]
  \centering
  \input{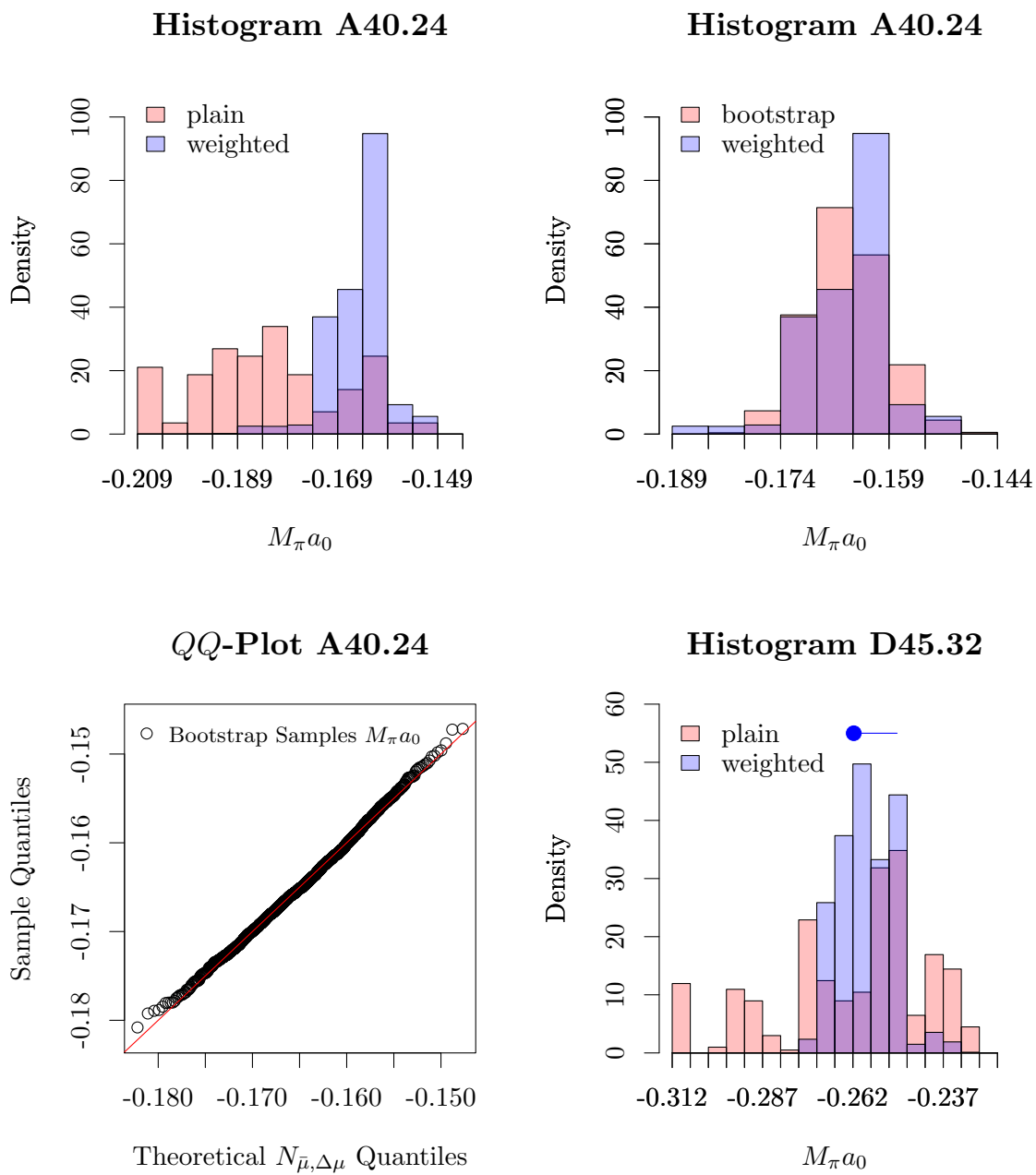}
  \caption{We show plain, weighted and bootstrap histograms for $M_\pi
  a_0$ and the ensembles A40.24 and D45.32, as explained in the
  text. In the lower left panel we show a QQ-plot for the
  bootstrap samples quantiles of $M_\pi a_0$ for A40.24.}
  \label{fig:hists}
\end{figure}

Using the weights $w_X$, we compute the weighted median over all
fit results on the original data to obtain our estimate for the expectation value $\langle
X\rangle$. The $68.54\%$ confidence interval of the weighted
distribution provides an (not necessarily symmetric) estimate for the
systematic uncertainty stemming from the different fit ranges. The
statistical error on $\langle X\rangle$ is computed from the $R$
bootstrap samples of the weighted median.

When derived quantities like $q\cot\delta$ are being determined, we
follow the same procedure, just that the weights are now given by the
products of the weights of the different contributing energy levels. 

The fit ranges are chosen such that for both the two point function and
the ratio at least five time slices are included in the fit. The two
point function is always fitted in an interval $[t_1^\pi,T/2]$ with
$t_1^\pi>6$. For the ratio we fitted in the interval $[t_1^{\pi\pi},
t_2^{\pi\pi}]$ with $t_1^{\pi\pi}\in\{11.5, 13.5, 15.5\}$ and
$t_2^{\pi\pi}\in\{22.5,21.5,20.5,19.5,18.5\}$ for $L=24,20$ and
$t_2^{\pi\pi}\in\{30.5,29.5,28.5,26.5,24.5\}$ for $L=32$. We note that
due to the weighting procedure described above we could have also
included smaller values for $t_1^\pi$ and $t_1^{\pi\pi}$ without
affecting the final result.

%As mentioned previously, we have determined the energy shift both from
%the ratio Eq.~\ref{eq:ratio} with only the four point function with
%$\vec p_1=\vec p_2=0$ entering and by solving a GEVP. The comparison
%shows that both procedures agree perfectly and, therefore, we restrict
%the presentation here to the analysis of the ratio Eq.~\ref{eq:ratio}.

\subsection{Systematic Effects}
\label{sec_sys_eff}

One of the main issues in this investigation is the question of
systematic uncertainties in our analysis procedure. In particular, we
have to consider possible contributions by the neutral pion which is
much lighter than the charged pion on our ensembles due to twisted
mass isospin breaking effects. Possible effects on the $I=2$ pion
scattering length extraction have been discussed in detail in
Ref.~\cite{Feng:2009ij}. 

In Ref.~\cite{Feng:2009ij} the authors were not able to find evidence
of neutral pion contributions within their errors. Even though we have
significantly smaller statistical errors, we still do not have any
evidence for these effects within the statistical uncertainties. In
the pseudo-scalar two point function the possible effect is an excited
state with mass $M_\pi + M_{\pi^0}$ due to the neutral pion having the
vacuum quantum numbers in twisted mass lattice
QCD~\cite{Shindler:2007vp}. This excited state could, however, never be
identified. In the four-point function
there can be effects from either close-by excited states at small
Euclidean times or thermal states at large Euclidean times. Again,
we do not see any evidence of those effects in our data.
 
However, the analysis procedure detailed above is designed such that
such effects should be covered by the systematic error we determine from the
weighted distribution. This systematic error mirrors possible
deviations from the theoretically expected curve and contributions by
excited states or pollutions at large Euclidean times. 

The neutral pion might also contribute to the exponential finite size
corrections in our data. As discussed below, for $M_\pi$ and $f_\pi$
we take these twisted mass specific effects into account as determined in
Ref.~\cite{Carrasco:2014cwa} from the data. For $q \cot(\delta_0)$ 
finite size corrections specific to twisted mass have not been computed
in $\chi$PT and, hence, we can only include the corrections computed
in continuum $\chi$PT in Ref.~\cite{Buchoff:2008hh}. 
However, these finite size corrections
computed in continuum $\chi$PT  have little
influence on our results. Hence, we do not expect large effects from
twisted mass specific finite size effects. 

This conclusion is further supported by the fact that we do not
observe large discretisation errors in the results for $M_\pi
a_0$. Any contribution from the neutral pion should show up as a 
$\mathcal{O}(a^2)$ lattice artefact. Of course, all these indications
are still not enough to finally exclude such systematic effects in our
results. But we conclude that within our uncertainties they are
negligible. 

\begin{table}[t!]
  \centering
  \begin{tabular*}{1.\textwidth}{@{\extracolsep{\fill}}lrrr}
    \hline\hline
    ens & $a\,\delta E$ & $aq\cot\delta_0$ & $M_\pi a_0$ \\
    \hline\hline
    A30.32  &  $0.0037(1)(^{+1}_{-2})$ & $-0.92(3)(^{+2}_{-5})$ & $-0.133(4)(^{+6}_{-4}) $  \\ 
    A40.32  &  $0.0033(1)(^{+1}_{-1})$ & $-0.90(3)(^{+3}_{-2})$ & $-0.155(5)(^{+4}_{-3}) $  \\ 
    A40.24  &  $0.0082(3)(^{+4}_{-1})$ & $-0.87(3)(^{+5}_{-1})$ & $-0.164(5)(^{+2}_{-8}) $  \\ 
    A40.20  &  $0.0179(5)(^{+2}_{-1})$ & $-0.71(2)(^{+1}_{-1})$ & $-0.202(4)(^{+1}_{-2}) $  \\ 
    A60.24  &  $0.0076(2)(^{+1}_{-1})$ & $-0.79(1)(^{+1}_{-1})$ & $-0.217(4)(^{+2}_{-3}) $  \\ 
    A80.24  &  $0.0071(1)(^{+0}_{-1})$ & $-0.75(1)(^{+0}_{-0})$ & $-0.262(3)(^{+1}_{-1}) $  \\ 
    A100.24 &  $0.0063(1)(^{+1}_{-1})$ & $-0.75(1)(^{+1}_{-1})$ & $-0.294(3)(^{+3}_{-1}) $  \\ 
    B55.32  &  $0.0039(1)(^{+1}_{-1})$ & $-0.71(2)(^{+1}_{-1})$ & $-0.219(5)(^{+3}_{-2}) $  \\ 
    D45.32  &  $0.0084(2)(^{+0}_{-5})$ & $-0.45(1)(^{+0}_{-2})$ & $-0.262(6)(^{+12}_{-1})$  \\ 
    B35.32  &  $0.0041(2)(^{+1}_{-1})$ & $-0.82(3)(^{+2}_{-2})$ & $-0.151(6)(^{+3}_{-3}) $  \\ 
    B85.24  &  $0.0085(1)(^{+0}_{-1})$ & $-0.66(1)(^{+0}_{-1})$ & $-0.292(3)(^{+4}_{-1}) $  \\ 
    \hline\hline
  \end{tabular*}
  \caption{$\delta E$, $q\cot\delta_0$ and $M_\pi a_0$ computed with total zero momentum.}
  \label{tab:deltaEfromL}
\end{table}

\subsection{L{\"u}scher formula to $\mathcal{O}(1/L^5)$}
\label{sec_1overL5}

We will discuss several procedures to determine scattering parameters
from the data. The first of which is to consider Eq.~\ref{eq:luscher1}
to the order $1/L^5$ and, hence,
neglecting the contribution from the effective range, $a_0$ can be
determined from $\delta E$, $M_\pi$  and $L$ by numerically solving
Eq.~\ref{eq:luscher1} for $a_0$. The corresponding results for $a_0$
in units of $M_\pi$ can be found in table~\ref{tab:deltaEfromL}. 

As an illustration of our analysis procedure we show example histograms 
in figure~\ref{fig:hists}. With plain
histograms we mean histograms of the unweighted results for different
fit ranges. In the weighted histograms the weights
have been applied according to Eq.~\ref{eq:weight}. And with bootstrap data we
denote the median of the weighted distribution evaluated on the
bootstrap samples. In the histograms we plot the densities of the
distribution. 

In the upper left panel the plain and the weighted histogram of $M_\pi
a_0$ determined from Eq.~\ref{eq:luscher1} on the A40.24 ensemble are
compared. The weighting leads to a well defined peak in the
histogram, which is representative for the findings on most ensembles. 

In the upper right panel a comparison 
of the weighted histogram and the histogram of the bootstrap samples
of $M_\pi a_0$ again for A40.24 is shown. That the distribution of the
bootstrap samples is 
approximately normal can be inferred from the lower left panel, where
we show the QQ-plot of the bootstrap sample quantiles versus the
theoretical quantiles of a standard normal distribution $N_{\bar \mu, \Delta
  \mu}$. Here $\bar \mu$ is the estimate of $M_\pi a_0$ and $\Delta
\mu$ its statistical error determined from the standard deviation over
the bootstrap samples. This comparison indicates that the systematic
and statistical uncertainties are approximately of the same size.
This finding is again representative for most of the ensembles. 

\begin{figure}[t]
  \centering
  \includegraphics[width=.7\linewidth]{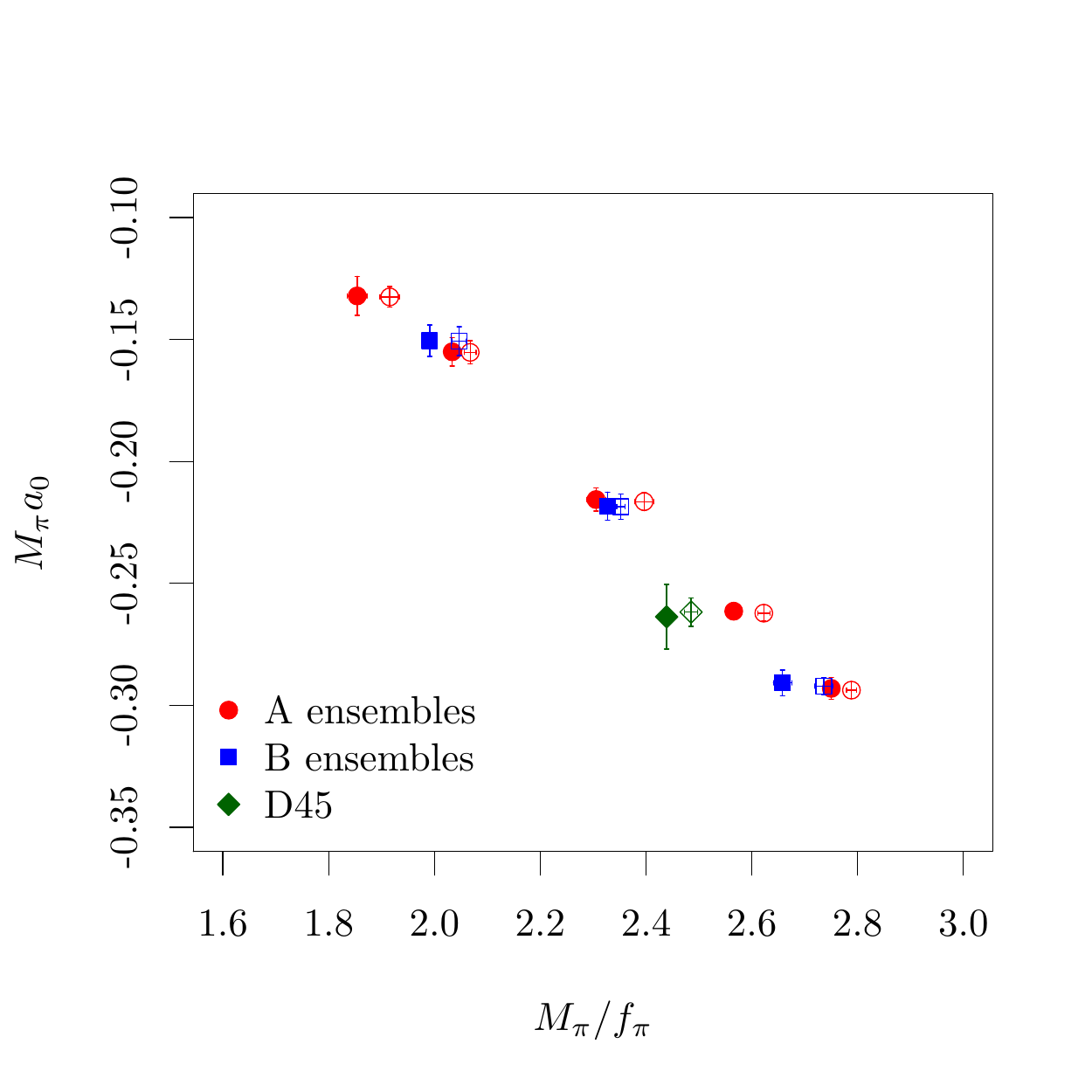}
  \caption{We show finite size corrected data for $M_\pi a_0$ as a
    function of $M_\pi/f_\pi$ with filled symbols compared to
    uncorrected data with open symbols. The corrected data include
    also the systematic uncertainty in order to allow for a comparison.}
      \label{fig:fs}
\end{figure}

In the lower right panel of figure~\ref{fig:hists} we show again the
plain versus the weighted histogram of $M_\pi a_0$, but this time for
ensemble D45. This ensemble shows the largest systematic uncertainty
of all the ensembles investigated here, 
which is also asymmetric.
The weighted median and the systematic
uncertainty which are indicated by the circle and the horizontal error
bar above the histogram, respectively, show this asymmetry clearly,
even if it is not easy to identify by eye.

For the successive analysis we also need to consider (exponentially
suppressed) finite size corrections to our data. For $M_\pi/f_\pi$ and
$M_\pi$ we use the results of Ref.~\cite{Carrasco:2014cwa} to correct
our data. The corresponding data for $M_\pi/f_\pi$ and the finite size
correction factors are summarised in table~\ref{tab:Mpi} in the
appendix. We remark
that A40.20 and D45.32 have not been considered in
Ref.~\cite{Carrasco:2014cwa}. For D45.32 we use the factors computed
in Ref.~\cite{Carrasco:2014cwa} for ensemble D20.48 with almost
identical $M_\pi L$-value. For A40.20 we do not need the finite size
corrections in the subsequent analyses.

For $a_0$ we apply the asymptotic finite size correction formula
Eq.~(31) from Ref.~\cite{Bedaque:2006yi}
\begin{equation}
  \label{eq:FSqcotdelta}
  \begin{split}
    \Delta(q\cot\delta) &= (q\cot\delta)_L - (q\cot\delta)_{L=\infty} \\
    &= \frac{-M_\pi}{\sqrt{2\pi}}\sum_{n=|\vec n|\neq
      0} c(n) \frac{e^{-nM_\pi L}}{\sqrt{nM_\pi L}}\left\{1-\frac{227}{24 n
        M_\pi L}+...\right\}\,,\\
  \end{split}
\end{equation}
which is valid close to threshold (small $q^2$) only. The $c(n)$ are
multiplicities for the $\vec n$ and can be found for instance in
Ref.~\cite{Bedaque:2006yi}. Note that at zero scattering momentum
$\lim_{q \to 0}q\cot\delta= \nicefrac{1}{a_0}$ holds which gives us
directly the finite size correction for the scattering length. 

For comparison we show the bare data for $M_\pi
a_0$ as a function of $M_\pi/f_\pi$ with open symbols and the
corresponding finite-size corrected data with filled symbols in figure~\ref{fig:fs}. 
From this plot it is visible that most of the finite-size effects in this analysis stem from the ratio
$M_\pi/f_\pi$. This is because the finite size corrections for $M_\pi$ and
$f_\pi$ are opposite in direction. One can also see the effect of
including the systematic uncertainties in the errors, which are
included for the finite size corrected data points, but not for the
uncorrected ones. This allows one to get an impression of their size
and for which ensembles they are actually relevant.

\subsection{L{\"u}scher Formula to $\mathcal{O}(1/L^6)$}
\label{sec_1overL6}

We have three A40 ensembles available, which differ only in their
spatial extends, $L=20,24$ and $L=32$, respectively. Here, we can apply
Eq.~\ref{eq:luscher1} in order to estimate the scattering length $a_0$
and the effective range $r_0$ from the $L$-dependence. It amounts to a
fit to the data for $\delta E$ for the three volumes with two fit
parameters. For $aM_\pi$ we used the values from the largest volume
ensemble A40.32. The result is summarised in the first column of
table~\ref{tab:deltaEA40}. 

\begin{table}[t!]
  \centering
  \begin{tabular*}{1.\textwidth}{@{\extracolsep{\fill}}lrrrr}
    \hline\hline
    & $1/L$-fit & $1/L$-fit & $q\cot\delta_0$-fit & $q\cot\delta_0$-fit\\
    \hline\hline
    $L$-values        & $32,24,20$ & $32,24$     & $32,24,20$ & $32,24$ \\
    $a_0/a$           & $-0.98(5)(^{+2}_{-16})$  & $-1.09(7)(^{+20}_{-9})$    & $-1.05(3)(^{+1}_{-11})$  & $-1.09(6)(^{+9}_{-9})$ \\
    $M_\pi a_0$       & $-0.138(6)(^{+2}_{-20})$ & $-0.154(10)(^{+28}_{-13})$ & $-0.149(4)(^{+1}_{-16})$ & $-0.154(8)(^{+12}_{-14})$ \\
    $r_0/a$           & $628(201)$               & $42(221)$                  & $147(22)(^{+30}_{-44})$  & $53(107)$ \\
    $M_\pi r_0$       & $98(28)$                 & $6(31)$                    & $21(3)(^{+4}_{-6})$      & $8(15)$ \\
    $M_\pi^2 a_0 r_0$ & $-87(23)$                & $-7(32)$                   & $-3.1(4)(^{+5}_{-9})$    & $-1(2)$ \\
    $\chi^2/\mathrm{dof}$ & $5.14$               & -                          & $0.79$                   & -\\
    \hline\hline
  \end{tabular*}
  \caption{A summary of our extractions of the scattering length
    $a_0$, and effective range $r_0$ for L{\"u}scher's $1/L$ expansion
    of the effective range expansion for two volume combinations
    each.} 
  \label{tab:deltaEA40}
\end{table}

\begin{figure}[t]
  \centering
  \subfigure[]{\includegraphics[width=.49\linewidth]{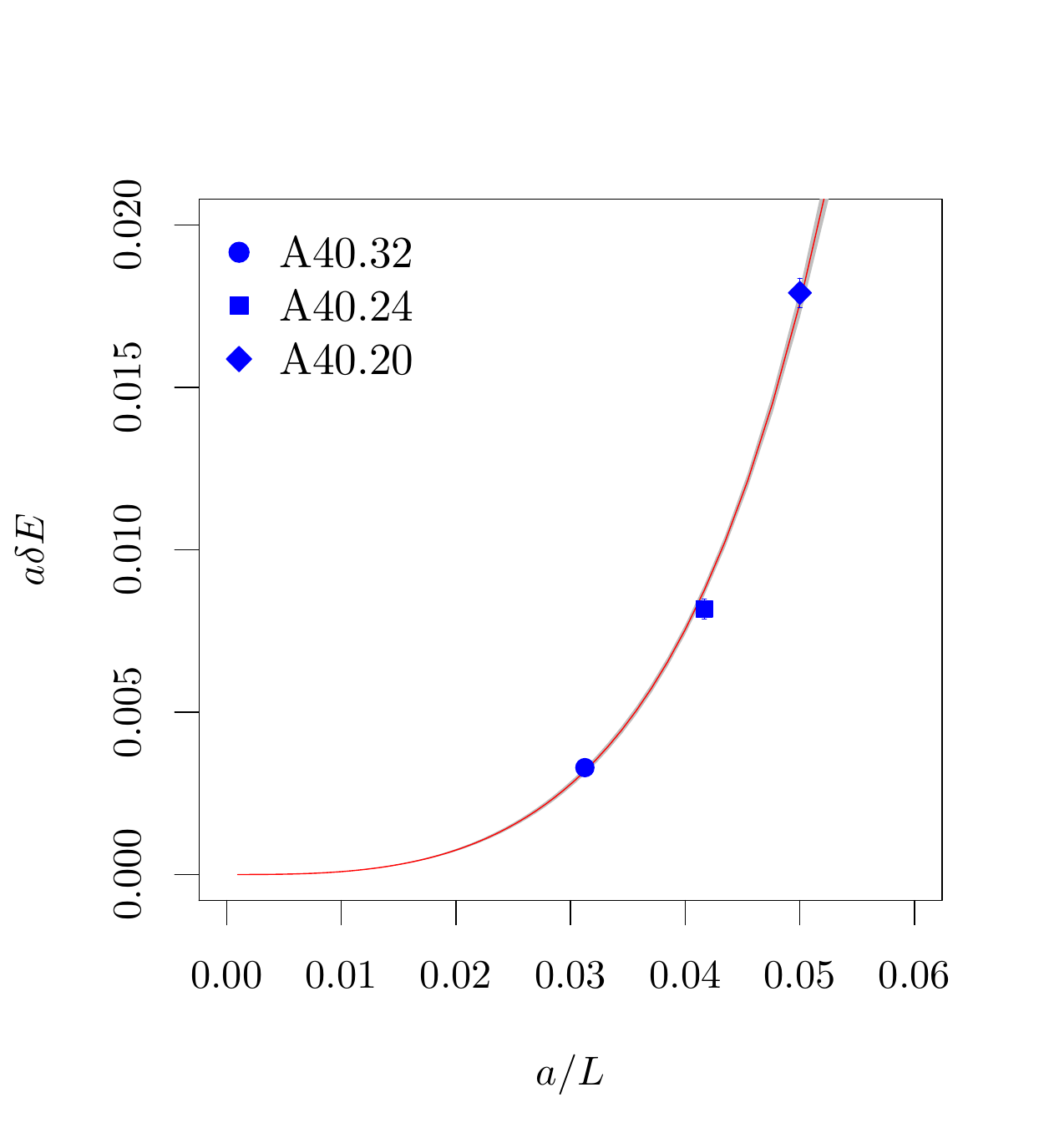}}
  \subfigure[]{\includegraphics[width=.49\linewidth]{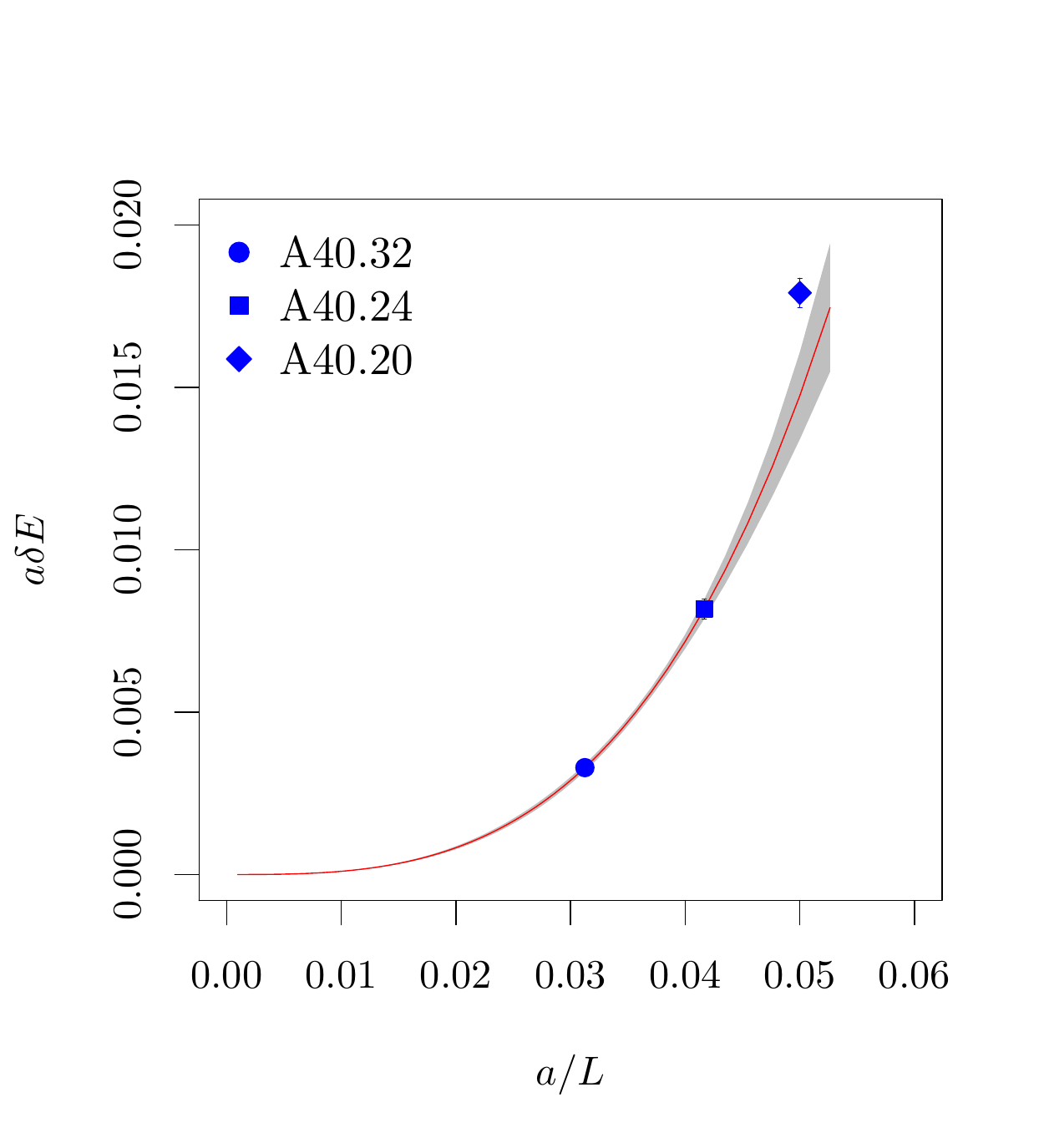}}
  \caption{(a) $a\delta E$ as a function of $a/L$ for the three A40
    ensembles and the best fit according to Eq.~\ref{eq:luscher1}. (b)
    the same like (a) but using A40.32 and A40.24 only.}
  \label{fig:deltaEA40}
\end{figure}

According to the $\chi^2$ value ($\mathrm{dof}=1$) this is not a good
fit. The data for $\delta E$ is shown together with the 
best fit and an error band in the left panel of
figure~\ref{fig:deltaEA40}. 
The result for $M_\pi a_0$ is lower than the one for the A40.32
ensemble alone (see table~\ref{tab:deltaEfromL}), but it deviates not more
than two $\sigma$. The effective range parameter $r_0$ is determined
by the fit only with large statistical uncertainties. 

It is likely that in particular A40.20 suffers still
from exponential finite size artefacts. Therefore, we repeat the
analysis with only A40.32 and A40.24 included. Now the fit basically
reproduces the result to $\mathcal{O}(1/L^5)$ of the A40.32 ensemble,
and there is no sensitivity to the effective range parameter. The fit
result is compiled in table~\ref{tab:deltaEA40} and shown in the right
panel of figure~\ref{fig:deltaEA40}. 

\subsection{Finite Range Expansion for $q\cot\delta_0$}
\label{sec_eff_range}

Instead of using the expansion in $1/L$ as defined in Eq.~\ref{eq:luscher1}, one may
also determine $q\cot\delta_0$ close to threshold directly and perform
the effective range expansion as follows:
\begin{equation}  \label{eq:ERexpansion}
  q\, \cot\delta_0\ =\ \frac{1}{a_0}\ +\ \frac{1}{2} r_0 q^2 + \mathcal{O}(q^4)\,,
\end{equation}
where $\delta_0$ is the s-wave phase shift, $a_0$ the corresponding
scattering length, $r_0$ the finite range parameter and $q$ the
scattering momentum. 
Using the three ensembles A40.32, A40.24 and A40.20 again,
we are able to obtain $q\cot\delta_0$ for three
values of the squared scattering momentum $q^2$. We then apply finite
size corrections using Eq.~\ref{eq:FSqcotdelta}.

\begin{figure}[t]
  \centering
  \subfigure[]{\includegraphics[width=.49\linewidth]{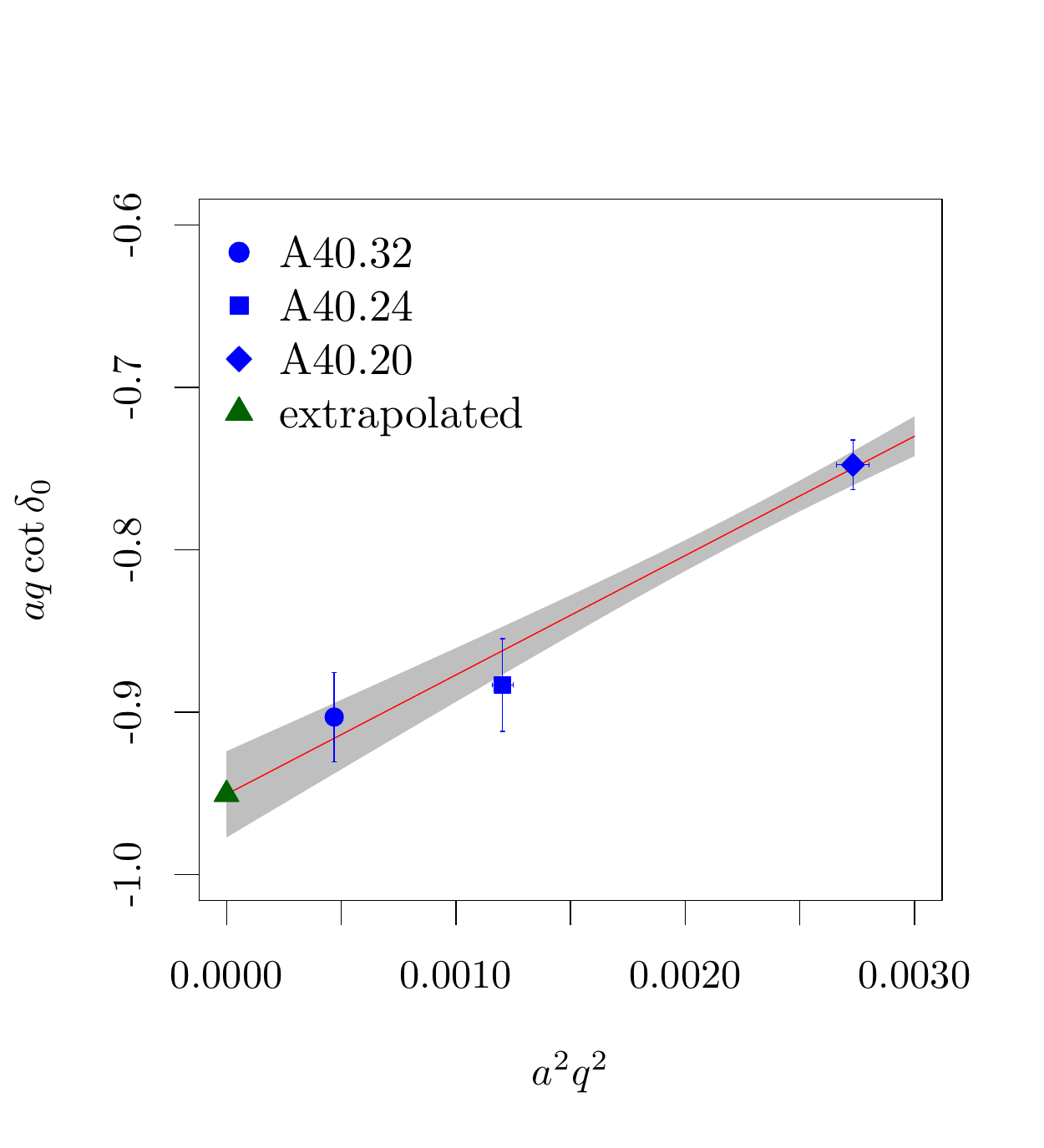}}
  \subfigure[]{\includegraphics[width=.49\linewidth]{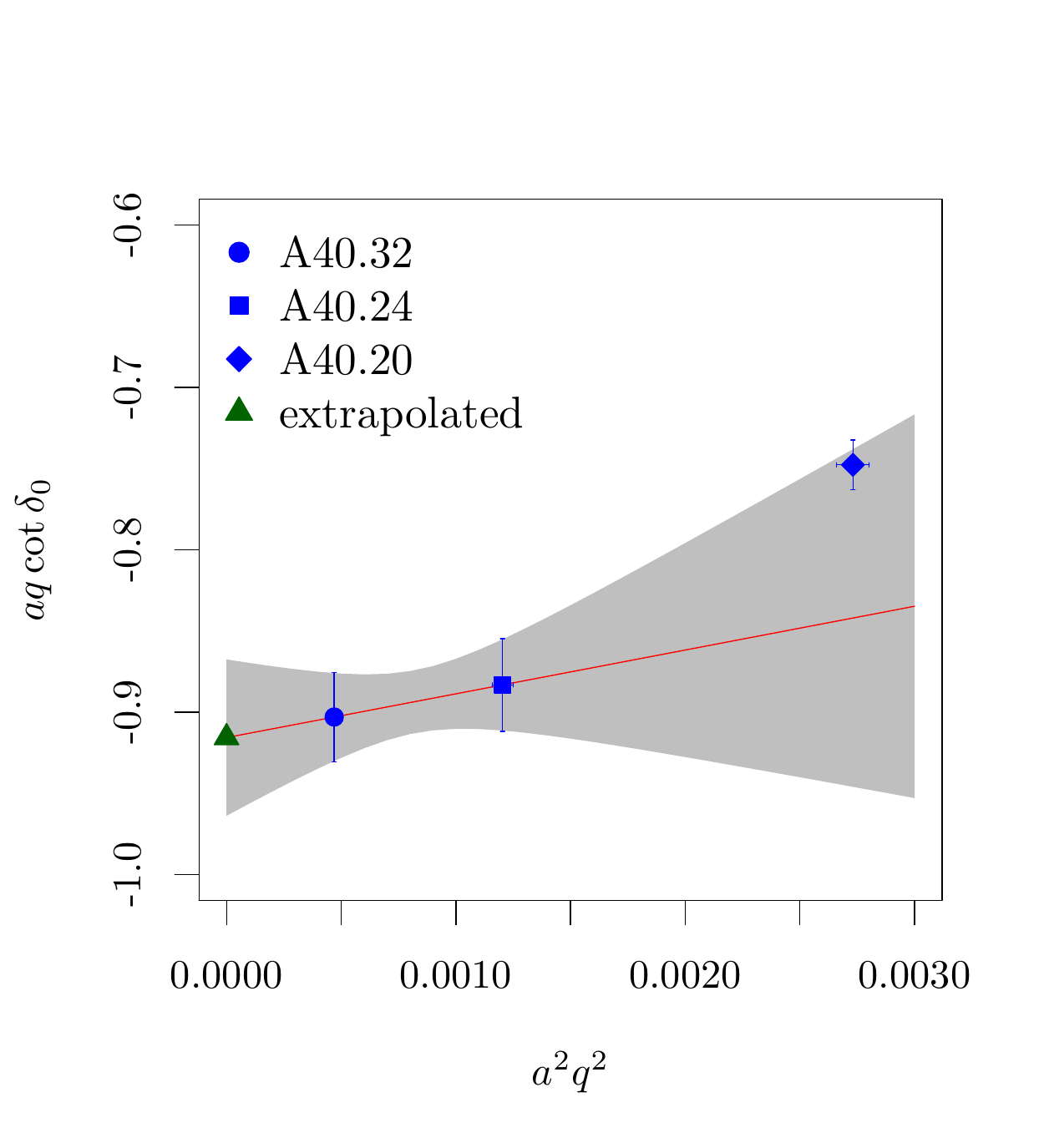}}
  \caption{$aq\cot\delta_0$ as a function of $a^2q^2$ for the three
    ensembles A40.32, A40.24 and A40.20 in (a) and for A40.32 and
    A40.24 only in (b). In addition the best fit of
    the effective range expansion to the data is shown as the line and
    the green triangle indicates the extrapolated value at $q^2=0$. The error band indicates the
    statistical uncertainty of the fit obtained from
    bootstrapping.} 
  \label{fig:ERqcotdelta}
\end{figure}

The corresponding data are shown in the left panel of figure~\ref{fig:ERqcotdelta} together
with the best fit to expression Eq.~\ref{eq:ERexpansion}.
The statistical error of the fit is
indicated by the grey band. In addition, the value extrapolated to $q^2=0$ is shown. 

The best fit result is summarised in the third column of
table~\ref{tab:deltaEA40}. In contrast to the fit of the $1/L$
expansion, the finite range
expansion provides a good description of $q\cot\delta_0$. The
$\chi^2$-value indicates a good fit, bearing in mind that there is
little freedom left. The statistical uncertainties
are smaller than the ones obtained from the $1/L$ expansion. In
particular, a statistically significant value for the effective range
parameter $r_0$ is obtained. 

Like in the previous case, where we used L{\"u}scher's direct method, we
also performed an extraction of $a_0$ and $r_0$ with only the largest
two volumes $L=24,32$ with the effective range formula. The result is
summarised in the last column of table~\ref{tab:deltaEA40} and the
data are shown in the right panel of
figure~\ref{fig:ERqcotdelta}. While the scattering length and
effective range are in good agreement with previous extractions, we
lose statistical significance for $r_0$ again. However, the error on
$a_0$ is only slightly bigger than found for the other three fits.
Note that the finite size corrections to $q\cot\delta_0$ do not change
the fit result significantly.

\begin{figure}[t]
  \centering
  \subfigure{\includegraphics[width=.48\linewidth]{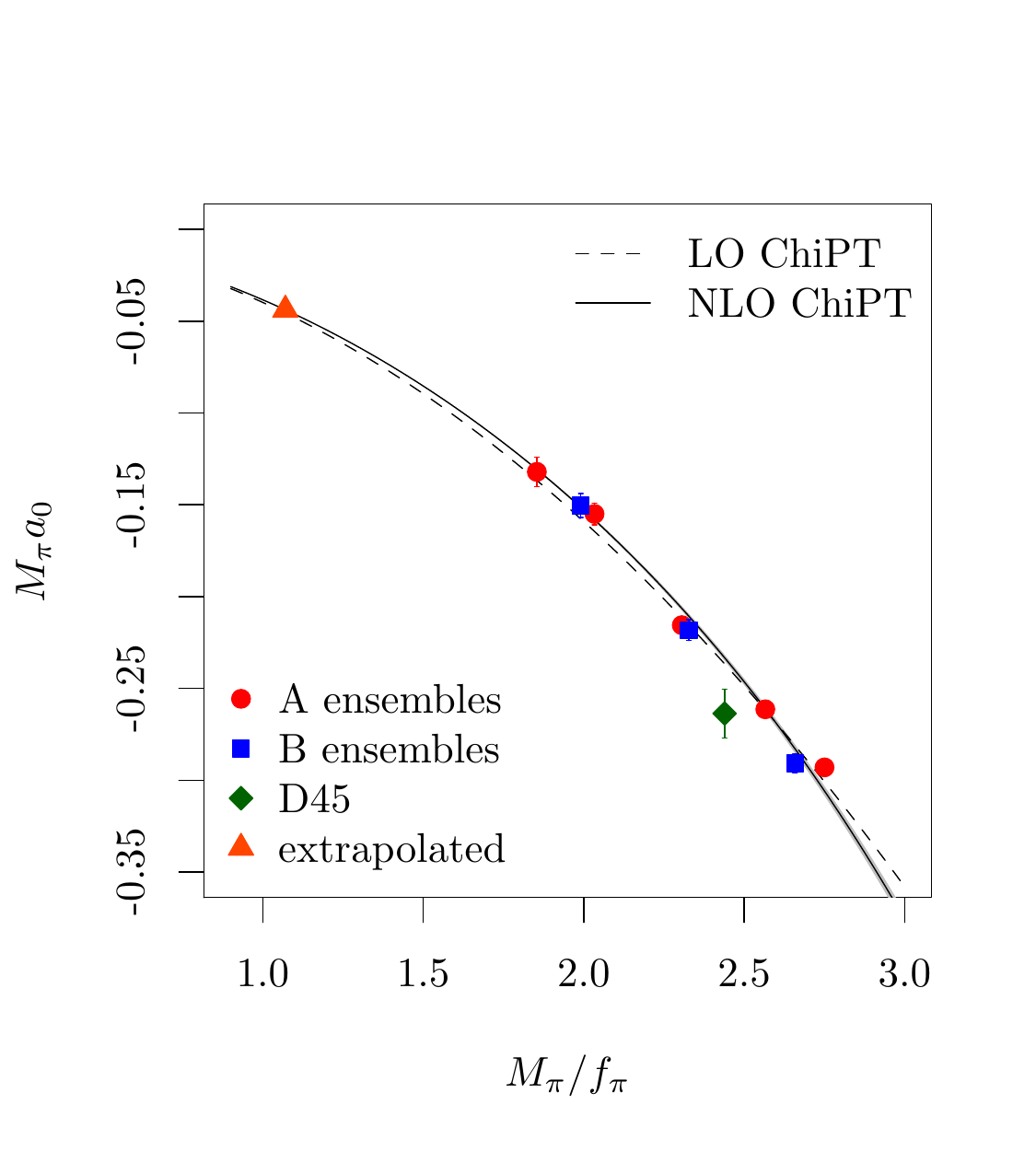}}
  \subfigure{\includegraphics[width=.48\linewidth]{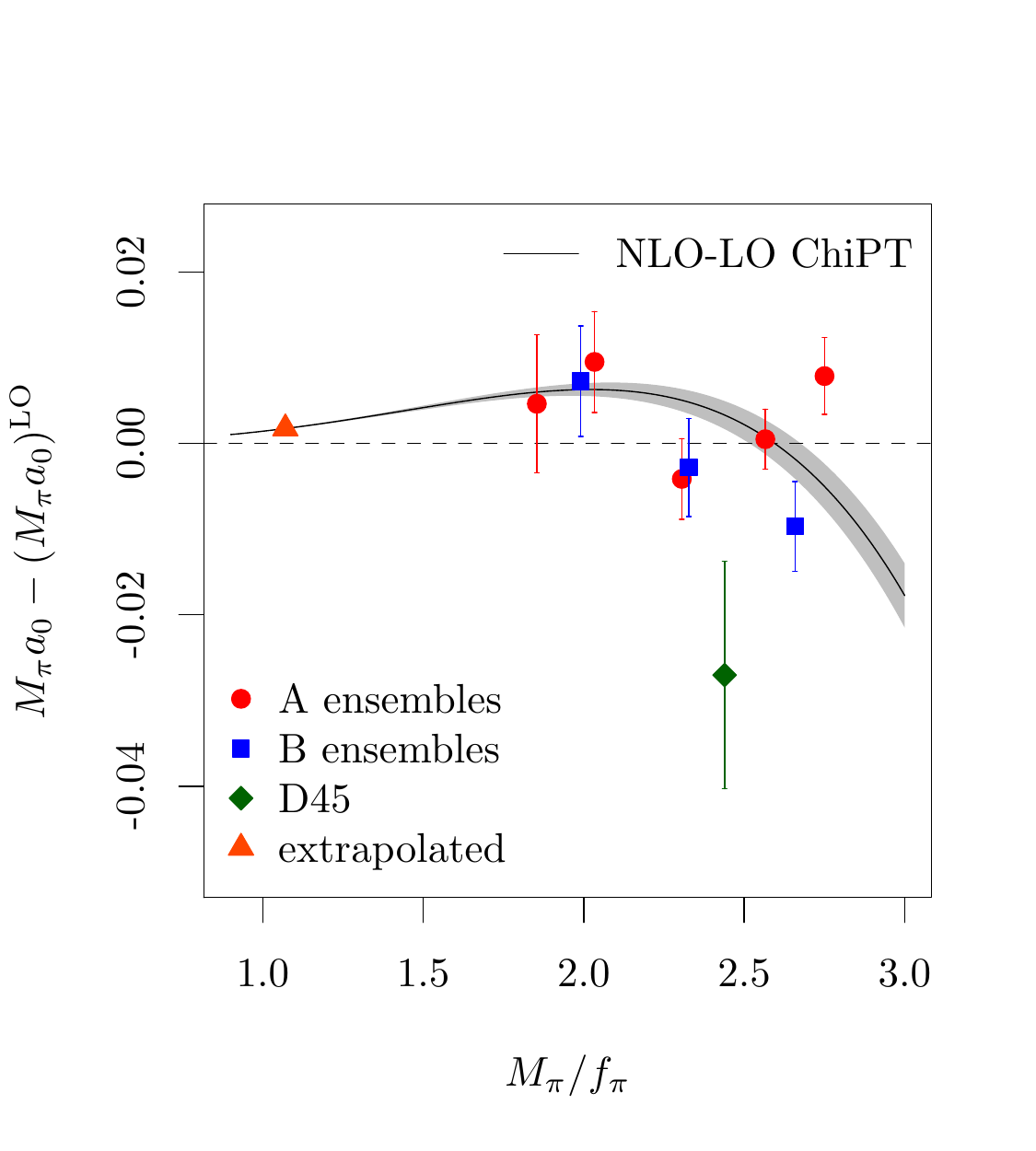}}
  \caption{Left: $M_\pi a_0$ as a function of $M_\pi/f_\pi$ determined
    from Eq.~\ref{eq:luscher1}  to $\mathcal{O}(L^{-5})$ with FS
    corrections for $M_\pi/f_\pi$ and $M_\pi a_0$. The A-ensembles do not include
    A40.24 and A40.20. We add the LO $\chi$PT prediction as the dashed
    line. The best fit to the data by the NLO $\chi$PT expression is
    shown as the solid line with error band. $M_\pi
    a_0|_{\mathrm{phys}}$ is plotted as the triangle. Right: the same
    but with LO $\chi$PT $(M_\pi a_0)^\mathrm{LO}$ subtracted.} 
  \label{fig:Mpia0ChPT}
\end{figure}

\subsection{Chiral Extrapolation}
\label{sec_chiral_extr}

Because we have determined the scattering parameters for bare quark masses
corresponding to larger than physical quark masses, we need to
extrapolate to the physical point. For the $I=2$ $\pi\pi$ scattering 
length $\chi$PT is particularly well suited, because it depends
on only one low energy constant at NLO.

As suggested in Refs.~\cite{Beane:2005rj,Beane:2007uh}, it is
convenient to write the $\chi$PT expression for $M_\pi a_0$ as a 
function of $M_\pi/f_\pi$, because all quantities are dimensionless
and no scale input is needed. The NLO $\chi$PT expression for $M_\pi
a_0$ written as a function of $(M_\pi/f_\pi)$ at a $\chi$PT
renormalisation scale $\mu_R=f_{\pi,\mathrm{phys}}$
reads~\cite{Beane:2005rj,Beane:2007uh} 
\begin{equation}
  \label{eq:chiptnlo}
  M_\pi a_0 = -\frac{M_\pi^2}{8\pi f_\pi^2}\left\{1 +
    \frac{M_\pi^2}{16\pi^2 f_\pi^2} \left[3\ln \frac{M_\pi^2}{f_\pi^2}
    - 1 - \ell_{\pi\pi}(\mu_R = f_{\pi,\mathrm{phys}})\right]\right\}
\end{equation}
with $\ell_{\pi\pi}$ related to the Gasser-Leutwyler
coefficients $\bar\ell_i$ as follows~\cite{Bijnens:1997vq}
\[
\ell_{\pi\pi}(\mu_R) = \frac{8}{3} \bar\ell_1 +
\frac{16}{3}\bar\ell_2 -\bar\ell_3 -4\bar\ell_4 +
3\ln\frac{M_{\pi,\mathrm{phys}}^2}{\mu_R^2}\,.
\]
Moreover, one can show in Twisted mass $\chi$PT that the leading
lattice artefacts to $M_\pi 
a_0$ are of order $\mathcal{O}(a^2M_\pi^2)$~\cite{Buchoff:2008hh}. At
NLO we, hence, consistently describe our data with the continuum
$\chi$PT formula provided above.

Since we are not (yet) able to determine the pion decay constant
$f_\pi$ with the sLapH approach, we use data
presented in Ref.~\cite{Michael:2013gka} for the ratio $M_\pi/f_\pi$ . The values for $M_\pi/f_\pi$ for all
ensembles are compiled in table~\ref{tab:Mpi} in the appendix. In this table we also give
the values for the finite size correction factors as determined in Ref.~\cite{Carrasco:2014cwa}. The finite size
corrections to $a_0$ are analytically computed using
Eq.~\ref{eq:FSqcotdelta} by setting $q\cot\delta_0\ =\ \nicefrac{1}{a_0}$.
For the finite size corrections of $M_\pi$ in $M_\pi a_0$ we also use the
factors compiled in table~\ref{tab:Mpi} as discussed above. 

The ratio $M_\pi/f_\pi$ can be determined with significantly smaller
uncertainty than $M_\pi a_0$. Therefore, we do not expect that the
missing statistical correlation between $M_\pi/f_\pi$ and $M_\pi a_0$
plays any role in the following analysis. 

For the fit we propagate the errors on $M_\pi/f_\pi$ and the finite
size corrections on $M_\pi$ and $f_\pi$ using resampling. For the
error on $M_\pi a_0$ we add the statistical and
systematic uncertainty of $M_\pi a_0$ and the statistical
uncertainty of the finite size correction factor $K_{M_\pi}$ in quadrature. In the $\chi^2$ minimisation
we take errors both on $M_\pi a_0$ and $M_\pi/f_{\pi}$ into account.
We only include the large volume A40.32 ensemble (and not A40.24 and
A40.20) in the fit.

\begin{table}[t]
  \centering
  \begin{tabular*}{.9\textwidth}{@{\extracolsep{\fill}}rrrrrr}
    \hline\hline
    $\ell_{\pi\pi}$ & $M_{\pi}a_0|_\mathrm{phys}$ & $\chi^2$ & $\mathrm{dof}$ & p-value & $M_\pi/f_\pi$-cut \\ 
    \hline\hline
    5.13(80) &   -0.0437(3) &    0.16     &  2 &  0.93  &  2.2 \\
    3.94(45) &   -0.0441(2) &    1.97     &  4 &  0.74  &  2.4 \\ 
    3.68(41) &   -0.0442(1) &    4.64     &  5 &  0.46  &  2.5 \\ 
    4.73(19) &   -0.0438(1) &    17.2     &  8 &  0.028 &  - \\ 
    \hline\hline
  \end{tabular*}
  \caption{Results from an NLO $\chi$PT fit to our data with different
    cuts in the upper fit-range.}
  \label{tab:ChPTfit}
\end{table}

In the left panel of figure~\ref{fig:Mpia0ChPT} we show the fit to all
the data (i.e. with no cut in $M_\pi/f_\pi$). The solid line
represents the best NLO $\chi$PT fit to our data, the dashed line the
LO parameter-free $\chi$PT prediction for $M_\pi a_0$. One observes
that the LO $\chi$PT prediction already describes the data
surprisingly well, and the NLO fit makes only a small correction to
the LO curve. We also show the value of $M_\pi a_0$ extrapolated to
the physical point. In the right panel of figure~\ref{fig:Mpia0ChPT}
we show the same, but with the LO $\chi$PT prediction $(M_\pi
a_0)^\mathrm{LO}$ subtracted.

The data for the A- and B-ensembles fall to a good approximation on a
single curve, whereas the only D-ensemble deviates 
slightly. While not inconsistent with a statistical
fluctuation, this might be due to the rather small physical
volume of the D45.32 ensemble. Another possible reason might be $a^2
M_\pi^2$ lattice artefacts together with higher order terms in
continuum $\chi$PT. 

\begin{figure}[t]
  \centering
  \subfigure{\includegraphics[width=.48\linewidth]{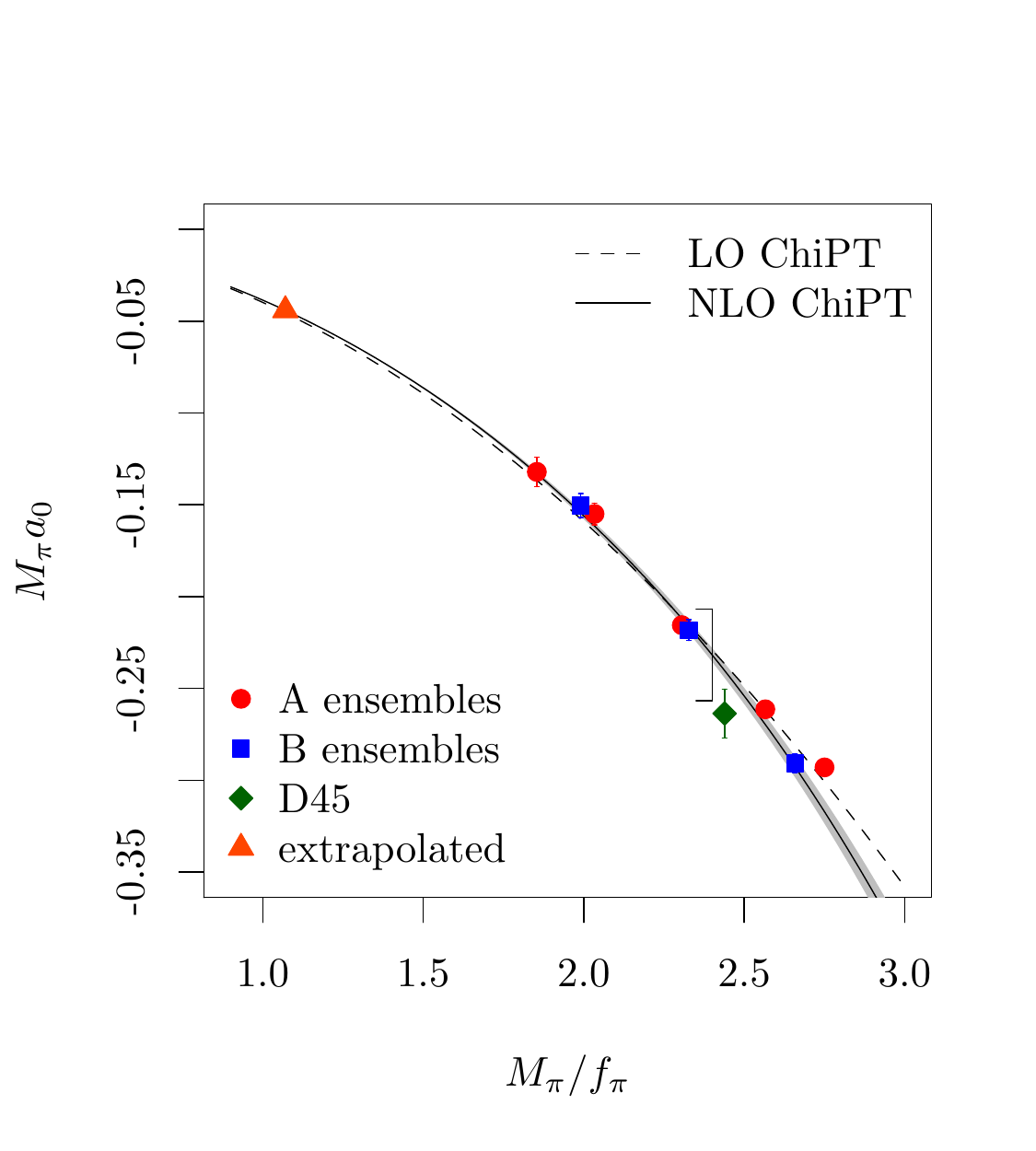}}
  \subfigure{\includegraphics[width=.48\linewidth]{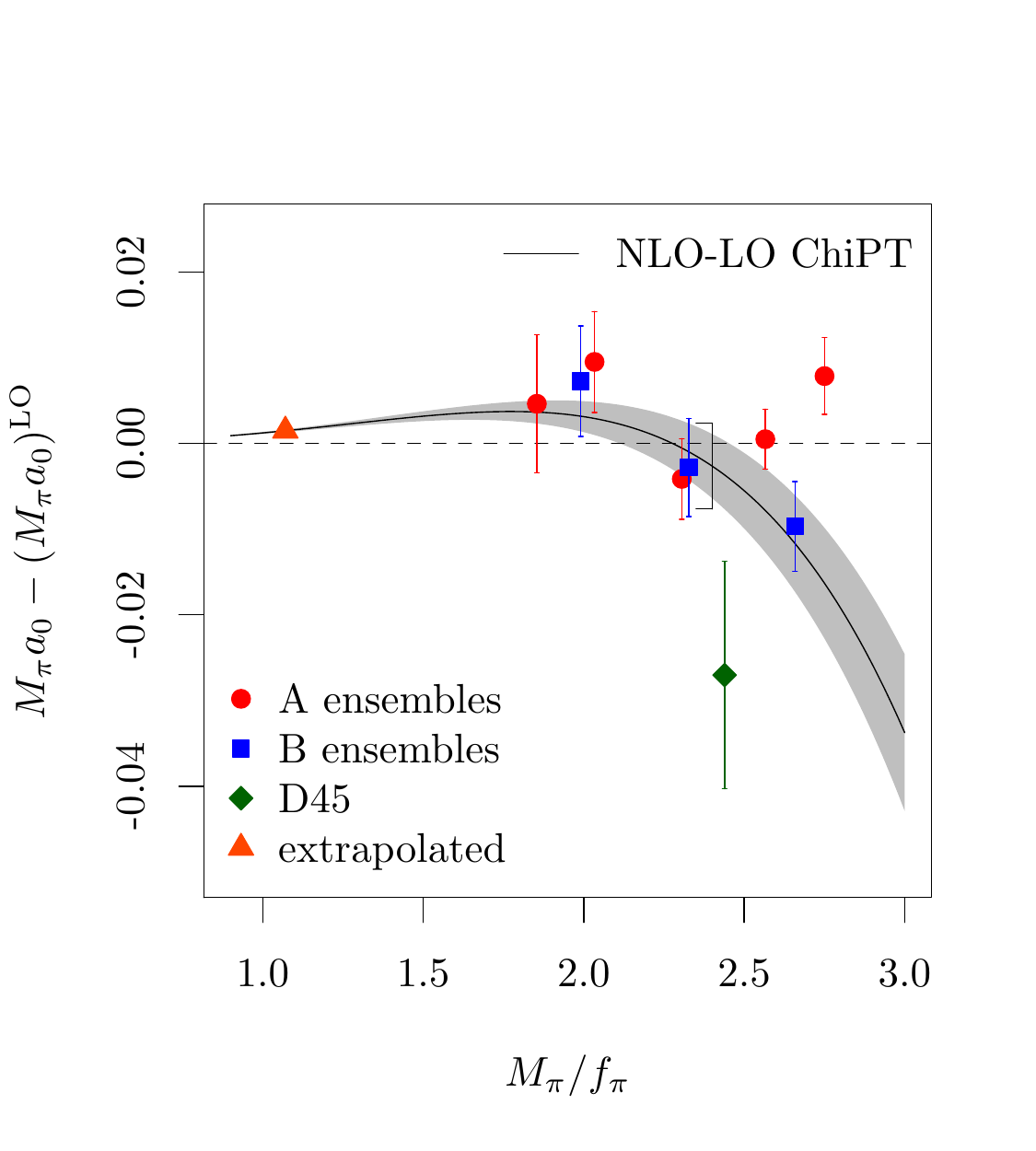}}
  \caption{Like figure~\ref{fig:Mpia0ChPT}, but for the fit including
    only data points with $M_\pi/f_\pi<2.4$ as indicated by the brackets.}
  \label{fig:ChPTcuts}
\end{figure}

We can explore this by restricting the fit range by applying an upper
cut in the $M_\pi/f_\pi$ values. For the case of the fit including
only data points with $M_\pi/f_\pi<2.4$ this is shown in
figure~\ref{fig:ChPTcuts}, which is otherwise identical to
figure~\ref{fig:Mpia0ChPT}. The cut value is indicated with brackets
in the plots.

The results of the fits to our data with different fit ranges are
summarised in table~\ref{tab:ChPTfit}. We observe that different fit
ranges do have little effect on the extrapolated value of $M_\pi
a_0|_\mathrm{phys}$. However, the fit quality improves with decreasing
upper fit range, with the best fit for the cuts $M_\pi/f_\pi<2.4$ and
$M_\pi/f_\pi<2.5$.
The only fit parameter $\ell_{\pi\pi}$ shows a
large variation, though the fitted values are mostly consistent within
errors. The reason for this large variation is visualised in the right
panels in figures~\ref{fig:Mpia0ChPT} and \ref{fig:ChPTcuts}. The
curvature is mainly driven by the points around $M_\pi/f_\pi=2$.
It is also visible that the fitted curves with cut still
describe the data reasonably well within errors also for $M_\pi/f_\pi$
values larger than the applied cut.

As a final result we quote the weighted average of the two fits with
cuts $M_\pi/f_\pi <2.4$ and $M_\pi/f_\pi<2.5$, respectively
\[
M_\pi a_0\ =\
-0.0442(2)_\mathrm{stat}(^{+4}_{-0})_\mathrm{sys}\,,\qquad
\ell_{\pi\pi}\ =\
3.79(0.61)_\mathrm{stat}(^{+1.34}_{-0.11})_\mathrm{sys}\,, 
\]
with the systematic error estimated from the maximal deviation of the
single results in table~\ref{tab:ChPTfit} to the finally quoted
one. The uncertainty stemming from the different fit-ranges of the fit
to the ratio data turns out to be sub-leading.

\section{Discussion}
\label{sec:discussion}

The results presented in the previous section require some
discussion. The first important point is that contributions from
thermal and excited states as well as the contributions from the
neutral pion can be excluded within the statistical uncertainty.
With the use of the ratio, Eq.~\ref{eq:ratio2}, thermal
states constant in time are minimised if not cancelled completely. In
section \ref{sec_1overL5}, e.g. figure~\ref{fig:hists}, we show
histograms created from fits to a large number of fit ranges. Even for
D45 where we see the largest systematic uncertainties stemming from
the different fit ranges, the systematic uncertainty is not much
bigger than the statistical uncertainty. This lets us conclude that we
cannot resolve any of the aforementioned systematic uncertainties within
our current statistical precision. 

Thanks to the three ensembles A40.32, A40.24 and A40.20 we
were able to extract scattering parameters
using three different methods: i) $1/L$ expansion,
Eq.~\ref{eq:luscher1}, to the order $1/L^5$ for every A40 ensemble
separately, ii) the same expansion but to the order $1/L^6$ and iii) a
direct fit of the effective range expansion to the values of
$q\cot\delta_0(q^2)$. Comparing the results in the
sections~\ref{sec_1overL5}, \ref{sec_1overL6} and \ref{sec_eff_range}
shows that all three methods give compatible results for $M_\pi a_0$,
apart from method i) for A40.20. The reason for the deviation of
method i) for A40.20 can be twofold: on the one hand $M_{\pi} L =
2.99$ might be too small and exponentially suppressed finite volume
effects cannot be ignored. On the other hand the physical volume 
might be too small so higher orders in $1/L$ in L{\"u}scher's formula and
higher orders in $q^2$ in the effective range expansion are
needed. Excluding A40.20 in methods ii) and iii) leads to smaller
$M_{\pi}a_0$ values, though still compatible within errors to the case
when including A40.20. These smaller values, however, are in better
agreement to method i) for both A40.32 and A40.24. From this fact
alone it is hard to deduce which kind of volume effect we are dealing
with here. That the physical volume of A40.20 might be too small is
supported by the fact that the D45 ensemble, which is the other
ensemble with small physical volume, gives a lower $M_{\pi}a_0$ than
expected although $M_{\pi} L = 3.87$ is in the range of the other
ensembles. The important consequence of this finding is that very
likely all other ensembles do not suffer from significant finite volume
effects. On every other ensemble than A40.20 and D45 the 
$M_{\pi} L$-value and the physical volume is at least as big as on
A40.24 which seems to be free from volume effects within our current
precision.

For methods ii) and iii) also the effective range $r_0$ can be
extracted from fits to the data. The best fit values vary
significantly and have mostly very large uncertainties. Only for one
fit (see table~\ref{tab:deltaEA40}) the $r_0$-value is significant. We
conclude that at the current level of available data and statistical
precision we are not able to reliably extract the effective range
parameter. 

The chiral extrapolation we present in section~\ref{sec_chiral_extr}
includes three values of the lattice spacing. The A- and B-ensembles
agree quite well, only at larger pion mass values small deviations
show up. Unfortunately, we have currently only one ensemble for the
smallest lattice spacing value available, namely D45.32. D45.32 has a
quite small physical volume, while $M_\pi L\approx 3.8$ which is
compatible to the other ensembles. Currently, we cannot finally
investigate the reason for this discrepancy. It could be a statistical
fluctuation, which is supported by the exceptionally large systematic
uncertainty for this ensemble, or a finite volume effect, as discussed
above. However, we can also not exclude a lattice artifact. We are
working on additional D-ensembles, which will allow us to investigate
this further.

\begin{table}[t!]
 \centering
 \begin{tabular*}{1.\textwidth}{@{\extracolsep{\fill}}llll}
   \hline\hline
   & $N_f$ & $M_\pi a_0$ & $\ell_{\pi\pi}$ \\
   \hline\hline
   LO $\chi$PT    &       & $-0.4438$ & \\
   CGL (2001)     &       & $-0.0444(10)$ & \\
   CP-PACS (2004) & 2     & $-0.0431(29)(-)$ & $-$\\
   NPLQCD (2006)  & 2+1   & $-0.0426(6)(3)$ & $3.3(6)(3)$ \\
   NPLQCD (2008)  & 2+1   & $-0.04330(42)_\mathrm{comb}$ & $6.3(1.2)_\mathrm{comb}$ \\ 
   ETM (2010)     & 2     & $-0.04385(28)(38)$ & $4.65(0.85)(1.07)$ \\
   ETM (2015)     & 2+1+1 & $-0.0442(2)(^{+4}_{-0})$ & $3.79(0.61)(^{+1.34}_{-0.11})$ \\
   Yagi (2011)    & 2     & $-0.04410(69)(18)$ & $5.8(1.2)(-)$ \\
   Fu (2013)      & 2+1   & $-0.04430(25)(40)$ & $3.27(77)(-)$ \\
   PACS-CS (2014) & 2+1   & $-0.04263(22)(41)$ & $-$ \\ 
   \hline\hline
 \end{tabular*}
 \caption{Compilation of results for $M_\pi a_0$ and $\ell_{\pi\pi}$
   including LO $\chi$PT, $\chi$PT and Roy
   equations~\cite{Colangelo:2001df} denoted as CGL,
   CP-PACS~\cite{Yamazaki:2004qb}, NPLQCD~(2006)~\cite{Beane:2005rj}, 
   NPLQCD~(2008)~\cite{Beane:2007xs}, ETM~(2013)~\cite{Feng:2009ij}, 
   this work denoted as ETM~(2015), Yagi et al.~\cite{Yagi:2011jn},
   Fu~\cite{Fu:2013ffa} and PACS-CS~\cite{Sasaki:2013vxa}.}
 \label{tab:comparison}
\end{table}

In table~\ref{tab:comparison} we provide a compilation of theoretical
determinations for $M_\pi a_0$ and $\ell_{\pi\pi}$ from the
literature: this includes the LO $\chi$PT prediction as well as
the value determined using $\chi$PT and Roy equations from
Ref.~\cite{Colangelo:2001df} denoted as CGL. For the lattice results
we have decided to include only direct determinations for which the
chiral extrapolation has been performed:
CP-PACS~\cite{Yamazaki:2004qb}, NPLQCD (2006)~\cite{Beane:2005rj},
NPLQCD (2008)~\cite{Beane:2007xs}, ETM (2013)~\cite{Feng:2009ij}, 
this work denoted as ETM (2015), Yagi et al.~\cite{Yagi:2011jn},
Fu~\cite{Fu:2013ffa} and PACS-CS~\cite{Sasaki:2013vxa}. We quote
statistical and -- where available -- systematic uncertainties
separately. For NPLQCD~(2008) there is only the combined statistical
and systematic uncertainty. 

\begin{figure}[t]
  \centering
  \includegraphics[width=\linewidth]{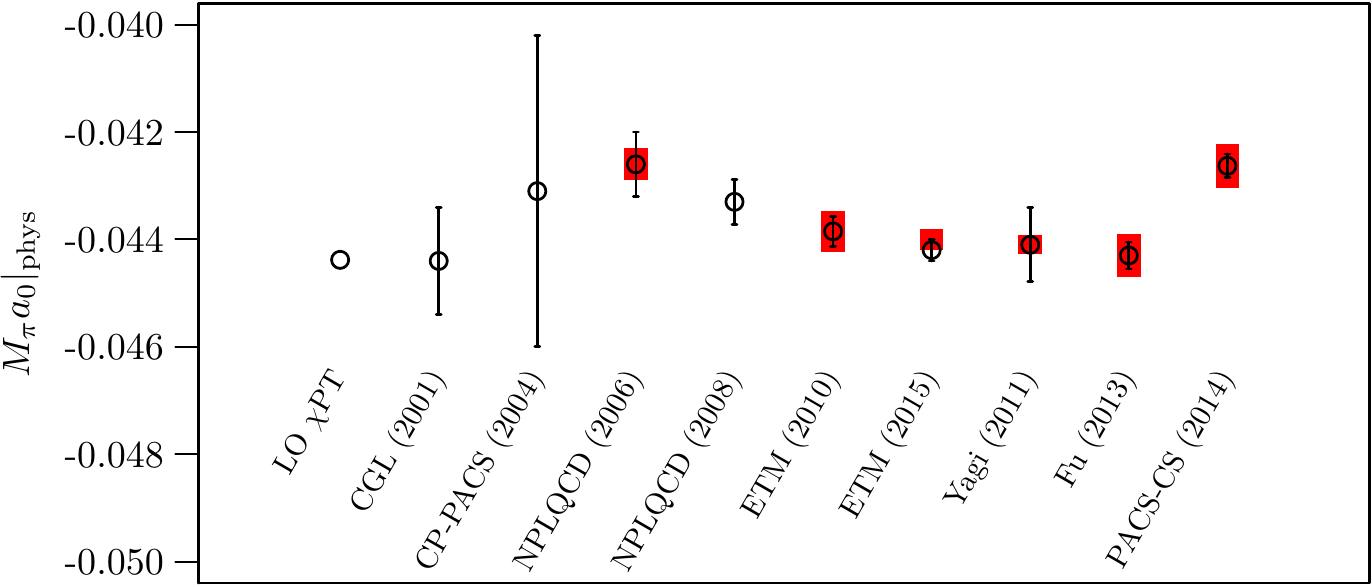}
  \caption{Comparison of the various predictions for $M_\pi
    a_0|_\mathrm{phys}$. The statistical error is indicated by
    the error bars on the points, the systematic uncertainty, where
    available, as coloured bars. Note that for NPLQCD~(2008) statistical and
    systematic errors are combined.}
  \label{fig:comparison}
\end{figure}

The CP-PACS and the here presented ETM~(2015) results are
based on three values of the lattice spacing, two have been use by the
authors of ETM~(2013) and by Fu. The others have used only one value
of the lattice spacing, however, NPLQCD~(2008) and PACS-CS have used
$\chi$PT to estimate discretisation effects. Finite size effects are
estimated in the works of NPLQCD~(2008), 
ETM~(2013), Yagi et al. and Fu using chiral perturbation theory. In
addition to the estimate from chiral perturbation theory we have
studied in this work several volumes to obtain an estimate of residual
finite volume effects.

As can be observed visually in figure~\ref{fig:comparison}, the values
for $M_\pi a_0$ show an overall good agreement. In particular, there
is no definite dependence on $N_f$. The exception are the high values
of NPLQCD~(2006) and PACS-CS. PACS-CS quotes in addition a small
uncertainty, which makes the deviation to our result statistically
significant. If the systematic uncertainty is added to the statistical
error, the results are compatible again. PACS-CS has studied the
smallest pion mass value around $170\ \mathrm{MeV}$ of all the works
considered in this comparison. However, they claim they cannot fit
this point with NLO $\chi$PT. Therefore, they had to include higher
order terms. More results at close to physical pion mass values might,
therefore, clarify whether this is a statistical or systematic effect
in the data, or whether the chiral extrapolation with NLO $\chi$PT is
misleading. 

The values for $\ell_{\pi\pi}$ show a large variation, which is,
however, covered by the large errors on this LEC.

\section{Summary}
\label{sec:summary}

In this paper, low-energy $\pi\pi$ scattering in the isospin $I=2$
channel is studied within L{\"u}scher's finite volume formalism in
lattice QCD. We use for the first time $N_f=2+1+1$ dynamical quark
flavours based on gauge
field configurations provided by ETMC. The list of ensembles covers
three values of the lattice spacing, several volumes and a large range
of pion mass values, see table~\ref{tab:setup}. We determine energy
shifts $\delta E=E_{\pi\pi} - 2M_\pi$ using the stochastic LapH method
and convert them in scattering length values applying L{\"u}scher's
formalism. 

We apply different, though closely related, methods to determine
scattering parameters and find compatible results for $M_\pi a_0$. The
finite range parameter $r_0$ cannot be determined with sufficient
certainty. 

Due to the use of ensembles with a broad range of pion masses we are
able to extrapolate $M_\pi a_0$ towards the physical pion mass point. 
After correcting for finite volume errors of the scattering length
$a_0$ we are able to make a rather smooth extrapolation towards the
physical pion mass value, the result of which is shown in
figure~\ref{fig:Mpia0ChPT}. We do not observe significant
discretisation effects between A and B-ensembles. The only D-ensemble
D45 shows a deviation which can be equally well explained by
discretisation effects, finite volume effects or by a statistical
fluctuation. To clarify this point, we are
generating data on more $D$ ensembles with smaller pion masses.

Our final result for $a^{I=2}_0$ is,  $M_\pi
a_0\ =\ -0.0442(2)_\mathrm{stat}(^{+4}_{-0})_\mathrm{sys}$. We have
compared our result to a list of other lattice determinations and
$\chi$PT combined with Roy equations and found mostly remarkably good
agreement.

The currently ongoing extension of the study presented in this paper
is $\pi\pi$ scattering with $I=1$, where the $\rho$ resonance is
present. With the perambulators  ready from the distillation process,
this is straightforward to do and first results are available.
For the pion-pion scattering, the channel $I=0$ is more complicated,
particularly for the twisted mass formulation due to isospin breaking
at finite lattice spacing values. Another currently ongoing extension is
to address low-energy scattering of other mesons: e.g. 
$\pi K$ and $K K$ scattering and more than two mesons.
Such techniques can also be
extended to charmed meson scattering processes relevant for the
recently discovered XYZ states.

\subsection*{Acknowledgements} 

We thank the
members of ETMC for the most enjoyable collaboration. The computer
time for this project was made available to us by the John von
Neumann-Institute for Computing (NIC) on the JUDGE and Juqueen
systems in J{\"u}lich. 
We thank U.-G.~Mei{\ss}ner for granting us 
access on JUDGE. We thank A.~Rusetsky for very useful discussions.
We thank K.~Ottnad for providing us with the data for
$M_\pi/f_\pi$ and S.~Simula for the estimates of the finite size
corrections to $M_\pi$ and $f_\pi$.
We thank C.~Michael for helpful comments on the manuscript.
This project was funded by the DFG as a project in
the Sino-German CRC110. C. Liu, J. Liu and J. Wang are supported in part by the
National Science Foundation of China (NSFC) under the project
 o.11335001. The open source software
packages tmLQCD~\cite{Jansen:2009xp}, Lemon~\cite{Deuzeman:2011wz} and
R~\cite{R:2005} have been used.

\bibliographystyle{h-physrev5}
\bibliography{bibliography}

\begin{thebibliography}{10}

\bibitem{Maiani:1990ca}
L.~Maiani and M.~Testa,
\newblock Phys.Lett. {\bf B245}, 585 (1990).
%%CITATION = PHLTA,B245,585;%%

\bibitem{Luscher:1985dn}
M.~L{\"u}scher,
\newblock Commun.Math.Phys. {\bf 104}, 177 (1986).
%%CITATION = CMPHA,104,177;%%

\bibitem{Luscher:1986pf}
M.~L{\"u}scher,
\newblock Commun.Math.Phys. {\bf 105}, 153 (1986).
%%CITATION = CMPHA,105,153;%%

\bibitem{Luscher:1990ck}
M.~L{\"uscher} and U.~Wolff,
\newblock Nucl.Phys. {\bf B339}, 222 (1990).

\bibitem{Luscher:1990ux}
M.~L{\"u}scher,
\newblock Nucl.Phys. {\bf B354}, 531 (1991).
%%CITATION = NUPHA,B354,531;%%

\bibitem{Rummukainen:1995vs}
K.~Rummukainen and S.~A. Gottlieb,
\newblock Nucl.Phys. {\bf B450}, 397 (1995), arXiv:hep-lat/9503028.
%%CITATION = HEP-LAT/9503028;%%

\bibitem{Kim:2005gf}
C.~Kim, C.~Sachrajda, and S.~R. Sharpe,
\newblock Nucl.Phys. {\bf B727}, 218 (2005), arXiv:hep-lat/0507006.
%%CITATION = HEP-LAT/0507006;%%

\bibitem{Feng:2011ah}
ETM, X.~Feng, K.~Jansen, and D.~B. Renner,
\newblock PoS {\bf LATTICE2010}, 104 (2010), arXiv:1104.0058.
%%CITATION = ARXIV:1104.0058;%%

\bibitem{Gockeler:2012yj}
M.~Gockeler {\em et~al.},
\newblock Phys.Rev. {\bf D86}, 094513 (2012), arXiv:1206.4141.
%%CITATION = ARXIV:1206.4141;%%

\bibitem{Li:2003jn}
X.~Li and C.~Liu,
\newblock Phys. Lett. {\bf B587}, 100 (2004), arXiv:hep-lat/0311035.
%%CITATION = HEP-LAT/0311035;%%

\bibitem{Feng:2004ua}
X.~Feng, X.~Li, and C.~Liu,
\newblock Phys. Rev. {\bf D70}, 014505 (2004), arXiv:hep-lat/0404001.
%%CITATION = HEP-LAT/0404001;%%

\bibitem{Detmold:2004qn}
W.~Detmold and M.~J. Savage,
\newblock Nucl.Phys. {\bf A743}, 170 (2004), arXiv:hep-lat/0403005.
%%CITATION = HEP-LAT/0403005;%%

\bibitem{Bedaque:2006yi}
P.~F. Bedaque, I.~Sato, and A.~Walker-Loud,
\newblock Phys.Rev. {\bf D73}, 074501 (2006), arXiv:hep-lat/0601033.
%%CITATION = HEP-LAT/0601033;%%

\bibitem{Doring:2011vk}
M.~Doring, U.-G. Meissner, E.~Oset, and A.~Rusetsky,
\newblock Eur.Phys.J. {\bf A47}, 139 (2011), arXiv:1107.3988.
%%CITATION = ARXIV:1107.3988;%%

\bibitem{Briceno:2013hya}
R.~A. Briceno, Z.~Davoudi, T.~C. Luu, and M.~J. Savage,
\newblock Phys.Rev. {\bf D89}, 074509 (2014), arXiv:1311.7686.
%%CITATION = ARXIV:1311.7686;%%

\bibitem{Agadjanov:2013kja}
D.~Agadjanov, U.-G. Mei?ner, and A.~Rusetsky,
\newblock JHEP {\bf 1401}, 103 (2014), arXiv:1310.7183.
%%CITATION = ARXIV:1310.7183;%%

\bibitem{He:2005ey}
S.~He, X.~Feng, and C.~Liu,
\newblock JHEP {\bf 0507}, 011 (2005), arXiv:hep-lat/0504019.
%%CITATION = HEP-LAT/0504019;%%

\bibitem{Liu:2005kr}
C.~Liu, X.~Feng, and S.~He,
\newblock Int.J.Mod.Phys. {\bf A21}, 847 (2006), arXiv:hep-lat/0508022.
%%CITATION = HEP-LAT/0508022;%%

\bibitem{Bernard:2010fp}
V.~Bernard, M.~Lage, U.-G. Meissner, and A.~Rusetsky,
\newblock JHEP {\bf 1101}, 019 (2011), arXiv:1010.6018.
%%CITATION = ARXIV:1010.6018;%%

\bibitem{Hansen:2012tf}
M.~T. Hansen and S.~R. Sharpe,
\newblock Phys.Rev. {\bf D86}, 016007 (2012), arXiv:1204.0826.
%%CITATION = ARXIV:1204.0826;%%

\bibitem{Briceno:2012yi}
R.~A. Briceno and Z.~Davoudi,
\newblock Phys.Rev. {\bf D88}, 094507 (2013), arXiv:1204.1110.
%%CITATION = ARXIV:1204.1110;%%

\bibitem{Guo:2012hv}
P.~Guo, J.~Dudek, R.~Edwards, and A.~P. Szczepaniak,
\newblock Phys.Rev. {\bf D88}, 014501 (2013), arXiv:1211.0929.
%%CITATION = ARXIV:1211.0929;%%

\bibitem{Roca:2012rx}
L.~Roca and E.~Oset,
\newblock Phys.Rev. {\bf D85}, 054507 (2012), arXiv:1201.0438.
%%CITATION = ARXIV:1201.0438;%%

\bibitem{Polejaeva:2012ut}
K.~Polejaeva and A.~Rusetsky,
\newblock Eur.Phys.J. {\bf A48}, 67 (2012), arXiv:1203.1241.
%%CITATION = ARXIV:1203.1241;%%

\bibitem{Briceno:2012rv}
R.~A. Briceno and Z.~Davoudi,
\newblock Phys.Rev. {\bf D87}, 094507 (2013), arXiv:1212.3398.
%%CITATION = ARXIV:1212.3398;%%

\bibitem{Hansen:2013dla}
M.~T. Hansen and S.~R. Sharpe,
\newblock (2013), arXiv:1311.4848.
%%CITATION = ARXIV:1311.4848;%%

\bibitem{Hansen:2015zga}
M.~T. Hansen and S.~R. Sharpe,
\newblock (2015), arXiv:1504.04248.
%%CITATION = ARXIV:1504.04248;%%

\bibitem{Hansen:2014lya}
M.~T. Hansen and S.~R. Sharpe,
\newblock (2014), arXiv:1409.7012.
%%CITATION = ARXIV:1409.7012;%%

\bibitem{Hansen:2014eka}
M.~T. Hansen and S.~R. Sharpe,
\newblock Phys.Rev. {\bf D90}, 116003 (2014), arXiv:1408.5933.
%%CITATION = ARXIV:1408.5933;%%

\bibitem{Beane:2007xs}
S.~R. Beane {\em et~al.},
\newblock Phys.Rev. {\bf D77}, 014505 (2008), arXiv:0706.3026.
%%CITATION = ARXIV:0706.3026;%%

\bibitem{Feng:2009ij}
X.~Feng, K.~Jansen, and D.~B. Renner,
\newblock Phys.Lett. {\bf B684}, 268 (2010), arXiv:0909.3255.
%%CITATION = ARXIV:0909.3255;%%

\bibitem{Dudek:2010ew}
J.~J. Dudek, R.~G. Edwards, M.~J. Peardon, D.~G. Richards, and C.~E. Thomas,
\newblock Phys.Rev. {\bf D83}, 071504 (2011), arXiv:1011.6352.
%%CITATION = ARXIV:1011.6352;%%

\bibitem{Beane:2011sc}
NPLQCD, S.~Beane {\em et~al.},
\newblock Phys.Rev. {\bf D85}, 034505 (2012), arXiv:1107.5023.
%%CITATION = ARXIV:1107.5023;%%

\bibitem{Dudek:2012gj}
J.~J. Dudek, R.~G. Edwards, and C.~E. Thomas,
\newblock Phys.Rev. {\bf D86}, 034031 (2012), arXiv:1203.6041.
%%CITATION = ARXIV:1203.6041;%%

\bibitem{Prelovsek:2014zga}
S.~Prelovsek,
\newblock PoS {\bf LATTICE2014}, 015 (2014), arXiv:1411.0405.
%%CITATION = ARXIV:1411.0405;%%

\bibitem{Frezzotti:2000nk}
ALPHA, R.~Frezzotti, P.~A. Grassi, S.~Sint, and P.~Weisz,
\newblock JHEP {\bf 08}, 058 (2001), hep-lat/0101001.
%%CITATION = HEP-LAT 0101001;%%

\bibitem{Frezzotti:2003ni}
R.~Frezzotti and G.~C. Rossi,
\newblock JHEP {\bf 08}, 007 (2004), hep-lat/0306014.
%%CITATION = HEP-LAT 0306014;%%

\bibitem{Baron:2010bv}
ETM, R.~Baron {\em et~al.},
\newblock JHEP {\bf 06}, 111 (2010), arXiv:1004.5284.
%%CITATION = 1004.5284;%%

\bibitem{Baron:2010th}
ETM, R.~Baron {\em et~al.},
\newblock Comput.Phys.Commun. {\bf 182}, 299 (2011), arXiv:1005.2042.
%%CITATION = ARXIV:1005.2042;%%

\bibitem{Frezzotti:2003xj}
R.~Frezzotti and G.~C. Rossi,
\newblock Nucl. Phys. Proc. Suppl. {\bf 128}, 193 (2004), hep-lat/0311008.
%%CITATION = HEP-LAT 0311008;%%

\bibitem{Chiarappa:2006ae}
T.~Chiarappa {\em et~al.},
\newblock Eur.Phys.J. {\bf C50}, 373 (2007), arXiv:hep-lat/0606011.
%%CITATION = HEP-LAT/0606011;%%

\bibitem{Peardon:2009gh}
Hadron Spectrum, M.~Peardon {\em et~al.},
\newblock Phys. Rev. {\bf D80}, 054506 (2009), arXiv:0905.2160.
%%CITATION = ARXIV:0905.2160;%%

\bibitem{Morningstar:2011ka}
C.~Morningstar {\em et~al.},
\newblock Phys.Rev. {\bf D83}, 114505 (2011), arXiv:1104.3870.
%%CITATION = ARXIV:1104.3870;%%

\bibitem{Hasenfratz:2001hp}
A.~Hasenfratz and F.~Knechtli,
\newblock Phys.Rev. {\bf D64}, 034504 (2001), arXiv:hep-lat/0103029.
%%CITATION = HEP-LAT/0103029;%%

\bibitem{Michael:1982gb}
C.~Michael and I.~Teasdale,
\newblock Nucl.Phys. {\bf B215}, 433 (1983).

\bibitem{Fukugita:1994ve}
M.~Fukugita, Y.~Kuramashi, M.~Okawa, H.~Mino, and A.~Ukawa,
\newblock Phys.Rev. {\bf D52}, 3003 (1995), arXiv:hep-lat/9501024.
%%CITATION = HEP-LAT/9501024;%%

\bibitem{Umeda:2007hy}
T.~Umeda,
\newblock Phys.Rev. {\bf D75}, 094502 (2007), arXiv:hep-lat/0701005.
%%CITATION = HEP-LAT/0701005;%%

\bibitem{Beane:2007qr}
S.~R. Beane, W.~Detmold, and M.~J. Savage,
\newblock Phys.Rev. {\bf D76}, 074507 (2007), arXiv:0707.1670.
%%CITATION = ARXIV:0707.1670;%%

\bibitem{Shindler:2007vp}
A.~Shindler,
\newblock Phys.Rept. {\bf 461}, 37 (2008), arXiv:0707.4093.
%%CITATION = ARXIV:0707.4093;%%

\bibitem{Carrasco:2014cwa}
ETM, N.~Carrasco {\em et~al.},
\newblock Nucl.Phys. {\bf B887}, 19 (2014), arXiv:1403.4504.
%%CITATION = ARXIV:1403.4504;%%

\bibitem{Buchoff:2008hh}
M.~I. Buchoff, J.-W. Chen, and A.~Walker-Loud,
\newblock Phys.Rev. {\bf D79}, 074503 (2009), arXiv:0810.2464.
%%CITATION = ARXIV:0810.2464;%%

\bibitem{Beane:2005rj}
NPLQCD, S.~R. Beane, P.~F. Bedaque, K.~Orginos, and M.~J. Savage,
\newblock Phys.Rev. {\bf D73}, 054503 (2006), arXiv:hep-lat/0506013.
%%CITATION = HEP-LAT/0506013;%%

\bibitem{Beane:2007uh}
NPLQCD, S.~R. Beane {\em et~al.},
\newblock Phys.Rev. {\bf D77}, 094507 (2008), arXiv:0709.1169.
%%CITATION = ARXIV:0709.1169;%%

\bibitem{Bijnens:1997vq}
J.~Bijnens, G.~Colangelo, G.~Ecker, J.~Gasser, and M.~Sainio,
\newblock Nucl.Phys. {\bf B508}, 263 (1997), arXiv:hep-ph/9707291.
%%CITATION = HEP-PH/9707291;%%

\bibitem{Michael:2013gka}
ETM, C.~Michael, K.~Ottnad, and C.~Urbach,
\newblock Phys.Rev.Lett. {\bf 111}, 181602 (2013), arXiv:1310.1207.
%%CITATION = ARXIV:1310.1207;%%

\bibitem{Colangelo:2001df}
G.~Colangelo, J.~Gasser, and H.~Leutwyler,
\newblock Nucl.Phys. {\bf B603}, 125 (2001), arXiv:hep-ph/0103088.
%%CITATION = HEP-PH/0103088;%%

\bibitem{Yamazaki:2004qb}
CP-PACS, T.~Yamazaki {\em et~al.},
\newblock Phys. Rev. {\bf D70}, 074513 (2004), arXiv:hep-lat/0402025.
%%CITATION = HEP-LAT/0402025;%%

\bibitem{Yagi:2011jn}
T.~Yagi, S.~Hashimoto, O.~Morimatsu, and M.~Ohtani,
\newblock (2011), arXiv:1108.2970.
%%CITATION = ARXIV:1108.2970;%%

\bibitem{Fu:2013ffa}
Z.~Fu,
\newblock Phys.Rev. {\bf D87}, 074501 (2013), arXiv:1303.0517.
%%CITATION = ARXIV:1303.0517;%%

\bibitem{Sasaki:2013vxa}
PACS-CS, K.~Sasaki, N.~Ishizuka, M.~Oka, and T.~Yamazaki,
\newblock Phys.Rev. {\bf D89}, 054502 (2014), arXiv:1311.7226.
%%CITATION = ARXIV:1311.7226;%%

\bibitem{Jansen:2009xp}
K.~Jansen and C.~Urbach,
\newblock Comput.Phys.Commun. {\bf 180}, 2717 (2009), arXiv:0905.3331.

\bibitem{Deuzeman:2011wz}
ETM, A.~Deuzeman, S.~Reker, and C.~Urbach,
\newblock (2011), arXiv:1106.4177.

\bibitem{R:2005}
{R Development Core Team},
\newblock {\em R: A language and environment for statistical computing},
\newblock R Foundation for Statistical Computing, Vienna, Austria, 2005,
\newblock {ISBN} 3-900051-07-0.

\end{thebibliography}

\newpage 
\begin{appendix}
  \section{Data Table}

\begin{table}[h!]
  \centering
  \begin{tabular*}{1.\textwidth}{@{\extracolsep{\fill}}lllll}
    \hline\hline
    & $aM_\pi$ & $M_\pi/f_\pi$ & $K_{M_\pi}$ & $K_{f_\pi}$ \\
    \hline\hline
    A30.32 & $0.1239(2)(^{+1}_{-1})$    & $1.915(10)$ & $1.0081(52)$ & $0.9757(61)$ \\
    A40.32 & $0.1415(2)(^{+1}_{-1})$    & $2.068(08)$ & $1.0039(28)$ & $0.9874(24)$ \\
    A40.24 & $0.1446(3)(^{+1}_{-1})$    & $2.202(13)$ & $1.0206(95)$ & $0.9406(84)$ \\
    A40.20 & $0.1474(6)(^{+1}_{-2})$    & NA          & NA           & NA \\
    A60.24 & $0.1733(3)(^{+5}_{-1})$    & $2.396(11)$ & $1.0099(49)$ & $0.9716(37)$ \\
    A80.24 & $0.1993(2)(^{+1}_{-0})$    & $2.623(07)$ & $1.0057(29)$ & $0.9839(22)$ \\
    A100.24& $0.2224(2)(^{+2}_{-0})$    & $2.788(07)$ & $1.0037(19)$ & $0.9900(15)$ \\
    B35.32 & $0.1249(2)(^{+1}_{-1})$    & $2.047(11)$ & $1.0069(32)$ & $0.9794(27)$ \\
    B55.32 & $0.1555(2)(^{+1}_{-1})$    & $2.352(07)$ & $1.0027(14)$ & $0.9920(10)$ \\
    B85.24 & $0.1933(3)(^{+0}_{-1})$    & $2.736(15)$ & $1.0083(28)$ & $0.9795(24)$ \\
    D45.32 & $0.1207(3)(^{+1}_{-1})$    & $2.485(12)$ & $1.0047(14)$ & $0.9860(13)$ \\ %FS taken from D20 with same Mpi*L
    \hline\hline
  \end{tabular*}
  \caption{Single pion energy levels, $M_\pi/f_\pi$ and the finite
    size correction factors $K_{M_\pi}$ and $K_{f_\pi}$ computed in
    Ref.~\cite{Carrasco:2014cwa} for $M_\pi$ and $f_\pi$,
    respectively.}
  \label{tab:Mpi}
\end{table}

\end{appendix}

\end{document}